%% file: wireless_label.tex
\documentclass[12pt, draftclsnofoot, onecolumn]{IEEEtran}
 \usepackage{amsmath,amssymb}
 \usepackage{subfigure}
 \usepackage{graphicx,graphics,color,psfrag}
 \usepackage{cite,balance}
 \usepackage{caption}
 \allowdisplaybreaks
 \usepackage{accents}
 \usepackage{amsthm}
 \usepackage{bm}
 \usepackage[english]{babel}
 \usepackage{multirow}
 \usepackage{enumerate}
 \usepackage{cases}
 \usepackage{stfloats}
 \usepackage{dsfont}
 \usepackage{color,soul}
 \usepackage{amsfonts}
 \usepackage{cite,graphicx,amsmath,amssymb}
 \usepackage{subfigure}
 \usepackage{fancyhdr}
 \usepackage{hhline}
 \usepackage{graphicx,graphics}
 \usepackage{array,color}
 \usepackage{amsmath}
 \usepackage{booktabs}
 \usepackage{algorithmic,algorithm}

\include{header}

\begin{document}
\setlength{\topskip}{-3pt}

\title{\huge Joint Annotator-and-Spectrum Allocation in Wireless Networks for Crowd Labelling}
\author{Xiaoyang Li, Guangxu Zhu, Kaiming Shen, Wei Yu, Yi Gong, and Kaibin Huang
\thanks{X. Li and Y. Gong are with the Department of Electrical and Electronic Engineering, Southern University of Science and Technology, Shenzhen, China. X. Li is also with the Department of Electrical and Electronic Engineering, The University of Hong Kong, Hong Kong.}
\thanks{G. Zhu is with the Shenzhen Research Institute of Big Data, Shenzhen, China.}
\thanks{K. Shen and W. Yu are with the Department of Electrical and Computer Engineering, University of Toronto, Canada.}
\thanks{K. Huang is with the Department of Electrical and Electronic Engineering, The University of Hong Kong, Hong Kong.} 
\thanks{Contact: K. Huang (huangkb@eee.hku.hk) or Y. Gong (gongy@sustech.edu.cn).} 
}
\maketitle

\vspace{-20mm}
\begin{abstract}
The massive sensing data generated by Internet-of-Things will provide fuel for ubiquitous \emph{artificial intelligence} (AI), automating the operations of our society ranging from transportation to healthcare. The realistic adoption of this technique however entails labelling of the enormous data prior to the training of AI models via supervised learning. To tackle this challenge, we explore a new perspective of \emph{wireless crowd labelling} that is capable of downloading data to many imperfect mobile annotators for repetition labelling by exploiting multicasting in wireless networks. In this cross-disciplinary area, the integration of the rate-distortion theory and the principle of repetition labelling for accuracy improvement gives rise to a new tradeoff between radio-and-annotator resources under a constraint on labelling accuracy. Building on the tradeoff and aiming at maximizing the labelling throughput, this work focuses on the joint optimization of encoding rate, annotator clustering, and sub-channel allocation, which results in an NP-hard integer programming problem. To devise an efficient solution approach, we establish an optimal sequential annotator-clustering scheme based on the order of decreasing signal-to-noise ratios. Thereby, the optimal solution can be found by an efficient tree search. Next, the solution is simplified by applying truncated channel inversion. Alternatively, the optimization problem can be recognized as a knapsack problem, which can be efficiently solved in pseudo-polynomial time by means of dynamic programming. In addition, exact polices are derived for the annotators constrained and spectrum constrained cases. Last, simulation results demonstrate the significant throughput gains based on the optimal solution compared with decoupled allocation of the two types of resources.
\end{abstract}


\section{Introduction}
The booming \emph{Internet of Things} (IoT) is expected to connect billions of devices to the Internet, resulting in a mass of global data \cite{poggi}. The availability of tremendous mobile data inspires researchers to envision ubiquitous computing and \emph{artificial intelligence} (AI) in wireless networks for automating operations of our society, ranging from transportation to healthcare \cite{zhu2018towards}. This leads to an emerging research area called \emph{edge machine learning}. One main focus in this area is on communication efficient techniques for acquiring and leveraging distributed data for AI model training \cite{zhu2018low,amiri2019machine,yang2018federated}. Given the enormity in the amount of collected raw data, labelling them is indispensable for training via supervised learning. However, this poses a challenge as labelling is a laborious and costly process. For instance, an average image set for deep learning consists hundreds of thousands of photos and it takes years to label thousands of such sets manually \cite{neuromation}. 

\subsection{Wireless Crowd Labelling}
Among others, the approach of crowd labelling, which exploits the wisdom of crowds, stands out for its low cost and high accuracy \cite{zhang2016learning}. With the emerging of crowd sensing platforms such as Amazon Mechanical Turk and CrowdFlower, the local labelling tasks can be distributed to and completed by ordinary Internet users \cite{ganti2011mobile}. A series of mechanisms are being developed to incentivize their participation including such as service \cite{zhang2015incentives}, money \cite{duan2012incentive}, and battery resource \cite{li2018wirelessly}. To guarantee the quality of crowd labelling, a series of algorithms have been developed to infer the ground true labels from multiple noisy labels. A simple but effective method is \emph{repetition labelling}, labelling the same object by multiple imperfect annotators, and combining their decisions by majority vote. Thereby, the accuracy of the inferred label is shown to increase with the number of noisy labels \cite{ipeirotis2014repeated}. Another vein of research, called machine learning based methods \cite{koller2009probabilistic}, focuses on using probabilistic graphical models and Bayesian inference to infer the true labels \cite{raykar2010learning,welinder2010multidimensional,whitehill2009whose,welinder2010online}. In view of prior work, communication overhead has been overlooked.

Statistics nowadays show that more than half of the users access Internet primarily through wireless terminal (including smartphones and tablet), and this number keeps growing. Consequently, downloading large datasets with high-dimensional samples to mobile annotators can exacerbate the congestion of wireless networks, which are already overloaded with many existing services ranging from mobile broadband to massive IoT for mission critical control. Hence, it is necessary to develop new wireless techniques to account for communication efficient crowd labelling across wireless networks. The uncharted area explored in this work is referred to as \emph{wireless crowd labelling}. Conventional wireless techniques attempt to achieve the goal of physical-layer rate maximization while wireless crowd labelling introduces computing elements into the processes e.g., label inference and a constraint on labelling accuracy. This makes the area cross-disciplinary involving interplay between techniques from wireless communication and machine learning. Consequently, a wide range of existing physical-layer techniques (ranging from multiple access to source encoding) need redesigning so as to maximize the communication efficiency of wireless crowd labelling. In this work, we consider data downloading via wireless multicasting for repetition labelling and focus on jointly managing the annotator-and-radio resources.

\subsection{Multicasting in Wireless Networks}
This work concentrates on the multimedia broadcasting network wherein a group of users may desire the same set of data. This gives rise to the scenario of multicasting in wireless networks, where users are clustered based on their interests and multiple data streams are transmitted to corresponding clusters simultaneously \cite{striccoli2018multicast}. The main issue for designing multicasting systems is that the links in a cluster support heterogeneous rates but they are used for transmitting the same data. There exist several ways for addressing the issue. The first is to optimize radio-resource allocation to increase all the link rates \cite{afolabi2012multicast}. This ensures the data is delivered reliably to all users in the cluster. Another way is to encode the data using superposition coding for supporting multi-rate multicasting \cite{won2009multicast}. As a result, the users can decode the same data but with different resolutions depending on their channel conditions. The last and the simplest strategy is to adapt the uniform transmission rate to the worst channel in the cluster to ensure reliable delivery to all at the cost of a reduced rate \cite{xu2011resource}.

With the rapid growth of multimedia services, multicasting is becoming increasingly important. This has been driving extensive efforts on designing different types of systems for supporting high-rate multicasting \cite{poularakis2016exploiting,zhou2017optimal,jiang2016multicast,tao2016content,sadeghi2017max,sharma2015genetic,xu2016energy,choi2015minimum,ding2017spectral}. For networks with content caching, the design of caching policies and scheduling for multicasting are jointly considered to minimize delay, power, and data-fetching costs \cite{poularakis2016exploiting,zhou2017optimal}. On the other hand, it is proposed that multicast delivery can be useful for reducing the energy consumption and thereby prolonging the lifetime of wireless multi-hop networks \cite{jiang2016multicast}. Transmission and resource allocation are studied for several popular communication systems including \emph{multiple-input-multiple-output} (MIMO) \cite{tao2016content,sadeghi2017max}, \emph{orthogonal frequency-division multiplexing} (OFDM) communication \cite{sharma2015genetic,xu2016energy}, and \emph{nonorthogonal multiple access} (NOMA) \cite{choi2015minimum,ding2017spectral}. For MIMO multicasting, researchers have studied the joint design of clustering and multicast beamforming \cite{tao2016content}, as well as the multicast precoding design accounting for user fairness \cite{sadeghi2017max}. For OFDM multicasting, an important topic is the joint (frequency) sub-channel and power allocation which has been investigated from the perspectives of rate maximization \cite{sharma2015genetic} and energy efficiency maximization \cite{xu2016energy}. Last, the coupling between users in NOMA system gives rise to new design challenges and solutions. In particular, the joint optimization of multicast beamforming and power allocation based on superposition coding is studied in \cite{choi2015minimum}. Moreover, a mixed multicast-unicast NOMA scheme is proposed in \cite{ding2017spectral}, which yields a significant improvement on the spectrum efficiency. 

Compared with the above conventional systems, multicasting for crowd labelling has its unique features and design challenges. First of all, users are now replaced with annotators and the purpose of multicast data is for repetition labelling but not for consumption. The transmission rate determines the level of data distortion and hence affects the accuracy of each annotator. Low accuracy can be overcome by crowd wisdom, namely repetition labelling with an increased number of annotators. Therefore, treating annotators as a type of resource, there exists a unique tradeoff between rates (or radio resource), annotator resource, and labelling accuracy. This motivates the current work to explore the new direction of joint radio-and-annotator resource allocation for enhancing labelling throughput under an accuracy constraint. 

\subsection{Contributions and Organization}
In this work, we consider an OFDM multicast system for wireless crowd labelling where an edge server multicasts multimedia objects to multiple clusters of annotators and in return collects the noisy labels for true-label inference. This work focuses on the joint optimization of encoding rate, annotator clustering, and sub-channel allocation to maximize the labelling throughput (i.e., the number of labelled objects) given the required accuracy. The problem is challenging for the following reasons. The link rates result in different quality of data samples received by annotators and thus the heterogeneous labelling accuracies. Then the first issue is how to cluster annotators so that repetition labelling of an object by each cluster can achieve the targeted accuracy. Next, given the rate-accuracy tradeoff, the labelling accuracies of clusters can be controlled via spectrum allocation and thus it need to be optimized. 

By deriving efficient solution approaches, this work makes the following contributions:
\begin{itemize}
\item {\bf Optimal Design with Fading Channels}: To deal with the challenging problem, we first advocate an optimal scheme for annotator clustering that can reduce the complexity. Specifically, clustering the annotators sequentially in a decreasing order of  \emph{signal-to-noise ratios} (SNRs) is proved to be globally optimal. This observation enables the construction of a tree graph for solving the original optimization problem. In other words, the height of the tree corresponds to the maximum labelling throughput under the optimal settings of annotator clustering and sub-channel allocation. Then an efficient algorithm is designed to search for the optimal solution on the graph based on branch-and-bound. 

\item {\bf Optimal Design with Channel Inversion}: To provide insight, we further derive a low-complexity solution and consider a special case with equalized channels (e.g., by inverse channel power control). Given identical channel gains, the optimal solution can be readily obtained from solving the classic knapsack problem that is solvable in pseudo-polynomial time by means of dynamic programming. Moreover, we provide an alternative approach in order to reduce complexity by node merging and graph truncation, i.e., a simplified type of the tree search. In terms of low complexity, the tree search approach is preferred to the knapsack one when the number of annotators is small, and the reverse is true if the number is large. 

\item {\bf Optimal Designs for Annotators/Spectrum Constrained Cases}: In this paper, the allocations of two resources, namely spectrum and annotators, are jointly optimized to maximize the labelling throughput. The general optimization problem can be either constrained by spectrum or annotators, which makes the problem numerically difficult.  By contrast, under the condition that either constraint is removed from the problem, much more efficient solution methods can be obtained. For the spectrum constrained case, the optimal policy is to encode each object at the lowest available rate and receptively label it using as many annotators as needed to meet the targeted accuracy. For the other case with annotators constrained, the opposite policy is optimal, where the highest encoding rate is adopted so as to minimize the number of annotators per cluster. 
\end{itemize}

\emph{Organization}: Section II introduces the system and mathematical models of its operations. The optimization problem is formulated in Section III. The solution approaches for the optimal design and that with truncated channel inversion are presented in Sections IV and V, respectively. The extension to the case of frequency selective fading is discussed in Section VI. Simulation results are provided in Section VII, followed by concluding remarks in Section VIII.

\begin{figure}[t]
\centering
\includegraphics[scale=0.8]{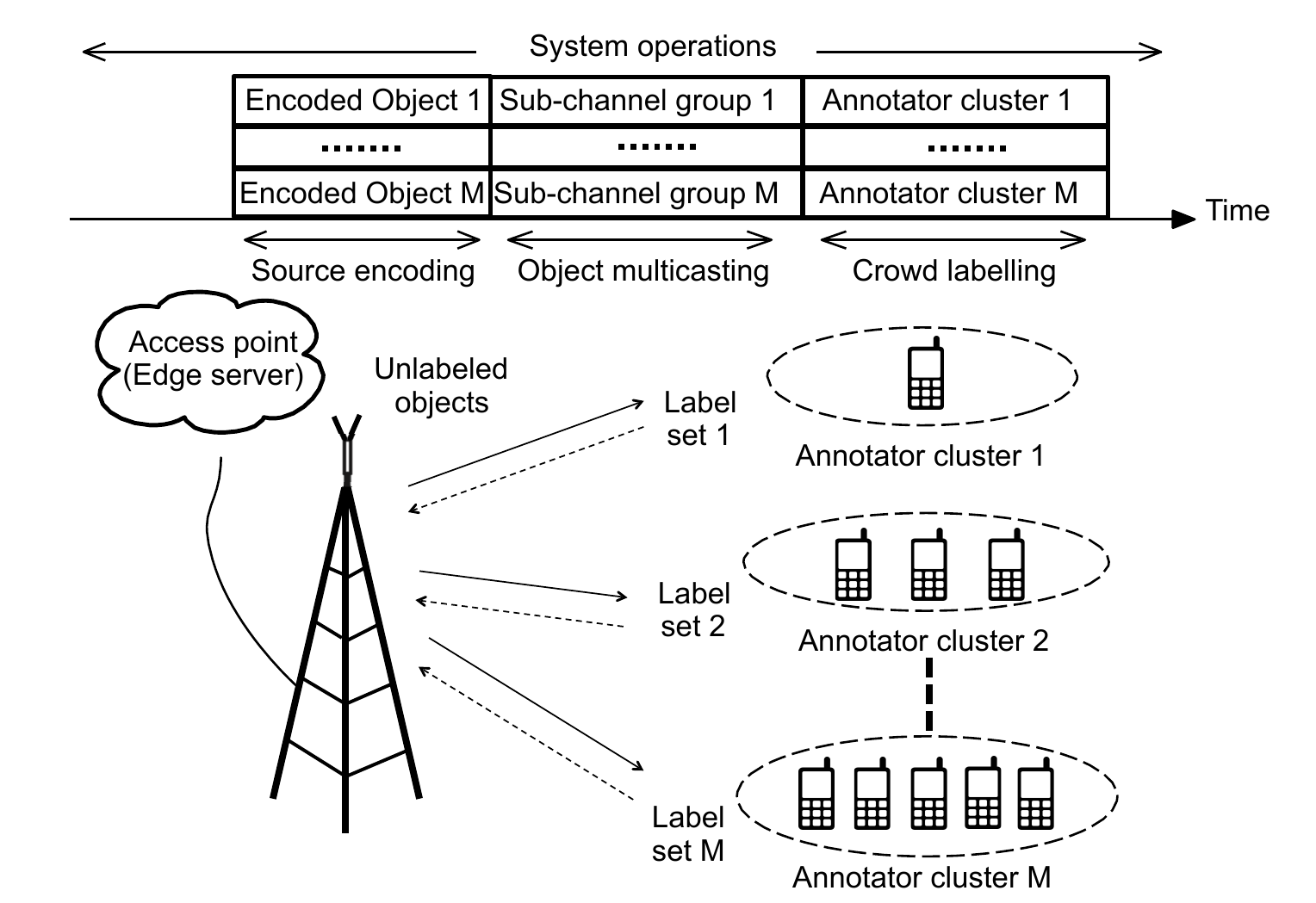}
\caption{Multicast wireless crowd labelling system.}
\label{FigSys}
\end{figure}

\section{System Model}
Consider a wireless labelling system as depicted in Fig.~\ref{FigSys}. In primary, the \emph{access point} (AP) has a sequence of raw data objects $(o_1,o_2,\dots,o_M)$ to be labelled by $K$ wireless annotators in a distributed fashion over a bandwidth $B$. We partition the set of annotators into $M$ non-overlapping subsets, whereas $\mathcal K_m \subseteq \{1,\dots,K\}$ is the subset of annotators labelling object $o_m$, $m=1,\ldots,M$. We partition the entire bandwidth into $M$ non-overlapping sub-channels, where a compressed version of the object $o_m$ is multicast in each sub-channel to be labeled by the set of annotators $\mathcal K_m$. The entire procedure consists of three stages---source encoding, object multicasting, and crowd labelling, which are specified separately in the rest of this section.

\subsection{Source Encoding Model}
Assume that every raw data object $o_m$ (e.g., video or image) is composed of $S$ bits. Prior to the wireless transmission, the AP first encodes object $o_m$ into $\hat o_m$ at rate $R_m$, so the size of $\hat o_m$ is $R_mS$ bits. In particular, we restrict the possible values of each $R_m$ to a discrete set $\{\lambda_1,\lambda_2,\ldots,\lambda_N\}$. The specific relation between $\hat o_m$ and $\lambda_n$ can be obtained from the rate-distortion tradeoff models in the existing literature, e.g., for video and images \cite{berger2003rate}. We use the binary variable $X^{(m)}_n\in\{0,1\}$ to indicate which encoding rate is adopted for object $o_m$, i.e., $X^{(m)}_n=1$ if $R_m=\lambda_n$ and $X^{(m)}_n=0$ otherwise. Obviously, we have

\begin{equation}
\sum^N_{n=1}X^{(m)}_n\le 1,~\forall m.
\end{equation}
In particular, object $o_m$ is nullified if $\sum^N_{n=1}X^{(m)}_n=0$, so it will not be transmitted or labeled in the subsequent stages. Since the radio resource is assigned to the objects sequentially, $\sum^N_{n=1}X^{(m)}_n=0$ only happens for the remaining objects when the resource is used up.
 
\subsection{Object Multicasting Model}
The coherence bandwidth $B$ is equally partitioned into $L$ sub-channels, $L_m$ of which are exclusively used for transmitting the encoded object $\hat{o}_m$ from the AP to the cluster of annotators in $\mathcal{K}_m$. (The non-coherence case with frequency-selective fading is discussed in Section IV.) With $T$ being the duration of wireless transmission, the maximum size of $\hat{o}_m$ that can be conveyed to annotator $k \in \mathcal{K}_m$ is
\begin{equation}\label{size}
Z_k = \frac{L_m B T}{L} \log_2(1+\gamma_k),
\end{equation}
where $\gamma_k$ represents the \emph{signal-to-noise ratio} (SNR) of the $k$th annotator. In order to guarantee successful reception of $\hat{o}_m$ at every annotators in $\mathcal{K}_m$, we must have
\begin{equation}\label{reception}
R_m S \leq \min_{k \in \mathcal{K}_m} Z_k.
\end{equation}

\subsection{Crowd Labelling Model}
Assume that all the annotators in $\mathcal{K}_m$ have received $\hat{o}_m$ successfully. For each annotator in $\mathcal{K}_m$, its \emph{labelling error probability} (LEP), denoted as $e_m$, is a monotonically decreasing function of the encoding rate $R_m$:
\begin{equation}\label{map}
e_m = f(R_m).
\end{equation}
The rationale is that the labelling task becomes more difficult when the distortion between $\hat{o}_m$ and $o_m$ increases (i.e., when the encoding rate decreases). The specific expression of function $f(\cdot)$ depends on the labelling algorithm and no particular form of $f(\cdot)$ is assumed in this paper.

Following the earlier work \cite{ipeirotis2014repeated}, we assume that the annotators in $\mathcal{K}_m$ are independent but exhibit the same LEP $e_m$. As a result, by applying the majority vote scheme \cite{ipeirotis2014repeated}, the decisions from all the annotators in $\mathcal{K}_m$ are combined into one decision, then achieve an overall error probability
\begin{equation}\label{RLEP}
P_m = \sum_{i=(|\mathcal{K}_m|+1)/2}^{|\mathcal{K}_m|} \binom{|\mathcal{K}_m|}{i} e_m^i (1-e_m)^{|\mathcal{K}_m|-i},
\end{equation}
which is referred to as the \emph{repetition LEP} (RLEP). Furthermore, by virtue of Stirling's formula \cite{mackay2003information}, the RLEP can be approximated as
\begin{equation}\label{approx}
\hat{P}_m = (4e_m(1-e_m))^{|\mathcal{K}_m|/2}.
\end{equation}

\section{Problem Formulation}
This work aims to find the optimal encoding rate $R_m$, annotator cluster $\mathcal{K}_m$, and sub-channel allocation $L_m$ for each of the $M$ objects, so as to maximize the number of objects that can be labelled with a RLEP below the target error probability $\theta>0$. The above joint optimization entails solving an integer programming problem:
\begin{subequations}
\begin{align}
\max_{\{X^{(m)}_n\},\{\mathcal{K}_m\},\{L_m\}} ~
& \sum_{m=1}^M \mathbf{1} \{\hat P_m \leq \theta\} \label{P1a}\\
\text{s.t.} ~
& R_{m} S \leq \min_{k \in \mathcal{K}_m} Z_k,~\forall~m,\label{P1b}\\
& \mathcal{K}_{m} \cap \mathcal{K}_{m'} = \emptyset,~\forall~m \neq m',\label{P1c}\\
\textbf{(P1)} \qquad \qquad \qquad & \mathcal{K}_m \subseteq \{1,\dots,K\},~\forall~m,\label{P1d}\\
& L_m \in \{0,1,\dots,L\},~\forall~m,\label{P1e}\\
& \sum_{m=1}^M L_m \leq L,\label{P1f}\\
& X^{(m)}_n \in \{0,1\},~\forall~m,\label{P1g}\\
&\sum^N_{n=1}X^{(m)}_n\le 1,\label{P1h}
\end{align}
\end{subequations}
where $\mathbf{1}\{\cdot\}$ represents an indicator function. The constraint \eqref{P1b} guarantees the successful reception of each object. The constraint \eqref{P1c} states that there is no annotator reuse for different clusters. The constraints \eqref{P1d} lists the possible clusters of annotators. The constraint \eqref{P1e} lists the possible sub-channel uses. The constraint \eqref{P1f} gives the sub-channel budget. The constraint \eqref{P1g} indicates which encoding rate is adopt for object $o_m$. The constraint \eqref{P1h} clarifies that each object can at most be encoded in one rate. The above problem is in essence an integer linear programming problem which is NP-hard in general \cite{schrijver2003combinatorial}.

\section{Optimal Design with Fading Channels}
To reduce the solution complexity, a simple but optimal annotator clustering strategy is derived in this section. Based on such strategy we can get rid of the variables $\mathcal K_m$ and $L_m$ in problem (P1), which can thus be solved by tree search. The structures of the optimal policy in two special cases are further studied, where either the spectrum or annotator resource is constrained.

\subsection{Optimal Annotator Clustering}
The Eq.~\eqref{P1b} implies that the transmission capacity from the AP to the cluster $\mathcal K_m$ is actually decided by the annotator with the worst SNR in the cluster, so we ought to group those annotators with close SNRs together. The proposed sequential clustering algorithm follows this idea: assume without loss of generality that $\gamma_1\ge\gamma_2\ge\ldots\ge\gamma_K$, then sequentially assign the annotators $\{1,\ldots,j_1\}$ to $\mathcal K_1$, the annotators $\{j_1+1,\ldots,j_2\}$ to $\mathcal K_2$ with $j_2>j_1$, and so forth. The proposition below shows that this simple algorithm is actually optimal, which is proved in Appendix~\ref{App:Optimal}.

\begin{proposition}[\emph{Optimal Annotator Clustering}]\label{Optimal}
The optimal clustering $\{\mathcal K^\star_m\}$ must be sequential.
\end{proposition}

In light of Proposition~\ref{Optimal}, optimizing the cluster set $\mathcal K_m$ amounts to deciding the last annotator $j_m$ (which is also the annotator with the worst SNR) in each cluster, so the maximum achievable rate for cluster $\mathcal K_m$ in Eq.~\eqref{P1b} can be rewritten as
\begin{equation}
\label{Z_cons}
\min_{k \in \mathcal{K}_m} Z_k = \frac{L_m B T}{L} \log_2(1+\gamma_{j_m}).
\end{equation}

As a result, the spectrum allocation and the annotation clustering can be optimally determined given the encoding rate decision $X^{(m)}_n$. First, if object $o_m$ is encoded at rate $\lambda_n$ (so $X^{(m)}_n=1$), the optimal number of annotators in cluster $\mathcal K_m$ is directly obtained from Eq.~\eqref{approx} and \eqref{P1a} as
\begin{equation}\label{annotator}
K_n = \l\lceil \frac{2\ln{\theta}}{\ln (4 f(\lambda_n)(1-f(\lambda_n)))} \r\rceil.
\end{equation}
As indicated by Eq.~\eqref{P1c} and \eqref{P1d}, the total annotator use should be no larger than $K$, i.e.,
\begin{equation}
\sum_{m=1}^M \sum_{n=1}^N K_n X_n^{(m)} \leq K.
\end{equation}

Due to the fact that there is no annotator reuse for different clusters as indicated by Eq.~\eqref{P1c}, the optimal $K_n$ further gives the index of the last annotator $j_m=j_{m-1}+K_n$. Since $j_1$ is uniquely determined by $X^{(1)}_n$, the optimal $j_m$ can be recursively obtained with $(X^{(1)}_n,\ldots,X^{(m)}_n)$. In other words, $j_m$ is a deterministic function of the sequence $(X^{(1)}_n,\ldots,X^{(m)}_n)$. After $j_m$ has been determined, the optimal $L_m$ can be readily obtained. With $X^{(m)}_n=1$, we aim to find the minimum possible $L_m$ that satisfies the Eq.~\eqref{P1b} and \eqref{P1e}, which has a closed-form solution:
\begin{equation}\label{subcarrier}
L_n^{(m)} = \l\lceil \frac{\lambda_n S L}{B T\log_2 (1+\gamma_{j_m})} \r\rceil.
\end{equation}
Recall that $j_m$ is a function of the sequence $(X^{(1)}_n,\ldots,X^{(m)}_n)$, and thus so is $L^{(m)}_n$. As indicated by Eq.~\eqref{P1f}, the total sub-channel use should be no larger than $L$, i.e.,
\begin{equation}
\sum_{m=1}^M \sum_{n=1}^N L_n^{(m)} X_n^{(m)} \leq L.
\end{equation}

Since $\hat P_m\le\theta$ and $\sum^N_{n=1}X^{(m)}_n = 1$ indicate that the object $o_m$ is successfully labelled, the objective in Eq.~\eqref{P1a} is equivalent to
\begin{equation}
\mathbf{1} \{\hat P_m\leq \theta\} = \mathbf{1} \{o_m~\text{is labelled}\} = \mathbf{1} \Bigg\{\sum_{n=1}^N X_n^{(m)} > 0\Bigg\}.
\end{equation}

Summarizing the above results, we arrive at the following reformulation of (P1):
\begin{subequations}
\begin{align}
\max_{\{X_n^{(m)}\}} \quad
& \sum_{m=1}^M \mathbf{1} \Bigg\{\sum_{n=1}^N X_n^{(m)} > 0\Bigg\} \label{P2a}\\
\text{s.t.} \quad
& \sum_{m=1}^M \sum_{n=1}^N K_n X_n^{(m)} \leq K,\label{P2b}\\
\textbf{(P2)} \qquad \qquad \qquad & \sum_{m=1}^M \sum_{n=1}^N L_n^{(m)} X_n^{(m)} \leq L,\label{P2c}\\
& \sum_{n=1}^N X_n^{(m)} \le 1,~ \forall~m,\label{P2d}\\
& X_n^{(m)}  \in \{0,1\},~\forall~m.\label{P2e}
\end{align}
\end{subequations}
Remarkably, the new problem involves the 0-1 variable $X^{(m)}_n$ alone. This simplification plays a key role in solving the integer programming problem via tree search, as discussed in the next sub-section.

\subsection{Tree Search Approach}
\begin{figure}[t]
\centering
\includegraphics[scale=0.55]{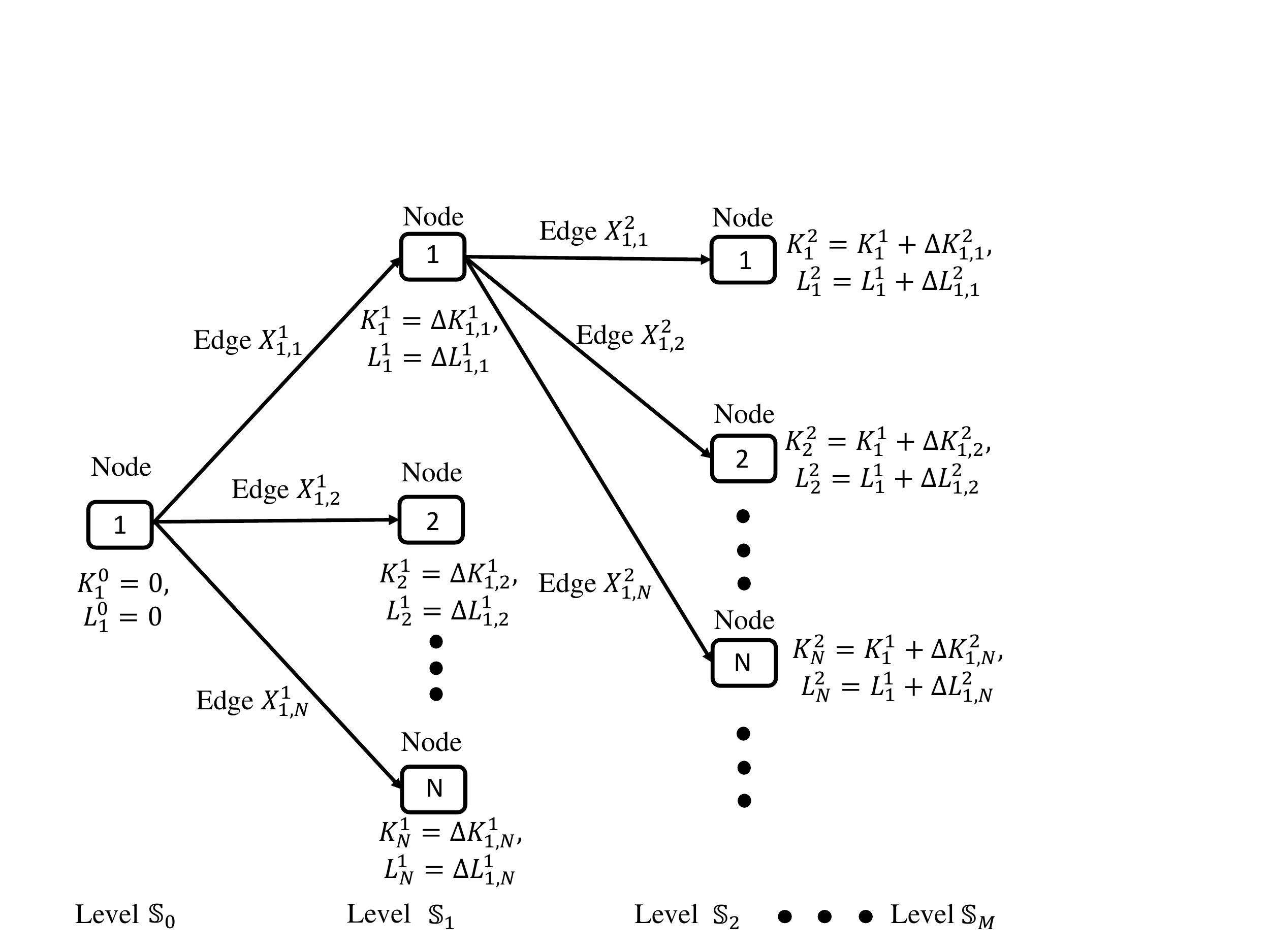}
\caption{The tree search graph for throughput maximization.}
\label{FigBBfading}
\end{figure}

In this section, we show that for throughput maximization, a search tree can be constructed to solve the problem (P2). Consider a tree graph displaying all combinations of annotator clustering and sub-channel allocation as shown in Fig.~\ref{FigBBfading}. In each level $\mathbb{S}_m$, there are $N$ possible clustering types (corresponds to $N$ encoding rates) for labelling the object $o_m$. The corresponding start and destination nodes are denoted as $s \in \mathbb{S}_{m-1}$ and $d \in \mathbb{S}_m$, respectively. The particular choice of each object can be viewed as an edge $X_{s,d}^{(m)}$ on the graph, where $X_{s,d}^{(m)} = 1$ indicates the edge is chosen, otherwise $X_{s,d}^{(m)} = 0$. As indicated by Eq.~\eqref{P2d}, each object can at most be labelled by one cluster, thus only one edge between two levels can be chosen, i.e., $\sum_{s \in \mathbb{S}_{m-1}} \sum_{d \in \mathbb{S}_m} X_{s,d}^{(m)} \leq 1,~\forall~m$. The connected edges forms a path, whose length gives the labelling throughput. Due to the limited resources, the path length cannot go to infinity, or equivalently the throughput is finite. Once the start and destination nodes are chosen, the clustering type for labelling the object is determined. Based on Eq.~\eqref{annotator}, the number of annotators can be expressed as
\begin{equation}\label{fadannotator}
\triangle K_{s,d}^{(m)} \equiv K_n = \l\lceil \frac{2\ln{\theta}}{\ln (4 f(\lambda_n)(1-f(\lambda_n)))} \r\rceil,
\end{equation} 
where $n = d - (s-1)N$. Based on Eq.~\eqref{subcarrier}, the sub-channel use can be expressed as
\begin{equation}\label{fadcarrier}
\triangle L_{s,d}^{(m)} = \l\lceil \frac{\lambda_n S L}{\log_2 (1+\gamma_{K_{d}^{(m)}}) B T} \r\rceil.
\end{equation} 
Correspondingly, the total annotator and sub-channel uses at node $d \in \mathbb{S}_m$ can be expressed as
\begin{equation}\label{Eq:IterK}
K_d^{(m)} = K_s^{(m-1)} + \triangle K_{s,d}^{(m)},
\end{equation} 
\begin{equation}\label{Eq:IterL}
L_d^{(m)} = L_s^{(m-1)} + \triangle L_{s,d}^{(m)},
\end{equation} 
where $K_s^{(0)} = 0$ and $L_s^{(0)} = 0$. The maximum labelling throughput is the height of the tree, i.e., the length of the longest path from the first level to the last level, which can be derived by an optimal \emph{joint annnotator and sub-channel allocation} (JASA) algorithm via branch-and-bound as shown in Algorithm~\ref{Al:BBfading}. A simple example is given below to illustrate how does the Algorithm~\ref{Al:BBfading} work.

\begin{algorithm}[tt]
\renewcommand{\algorithmicrequire}{\textbf{Input:}}
\renewcommand{\algorithmicensure}{\textbf{Output:}}
\caption{Throughput Maximization with Fading Channels.}
\label{Al:BBfading}
\begin{algorithmic}[1]
\REQUIRE number of objects to be labelled $M$, annotator and sub-channel budget $K$ and $L$, annotator and sub-channel uses $\triangle K_{s,d}^{(m)}$ and $\triangle L_{s,d}^{(m)}$
\ENSURE the maximum throughput $M^*$, and the optimal solution
\STATE Initialize $m=0$, $K_d^{(0)} = 0$ and $L_d^{(0)} = 0$
\STATE \textbf{Loop} $m = m+1$
\STATE \qquad \textbf{For} $d = 1:1:|\mathbb{S}_m|$
\STATE \qquad \qquad Create the $d$-th node with $K_d^{(m)} = K_{s}^{(m-1)} + \triangle K_{s,d}^{(m)}$ and $L_d^{(m)} = L_{s}^{(m-1)} + \triangle L_{s,d}^{(m)}$ 
\STATE \qquad \qquad \textbf{If} $K_d^{(m)}>K$ or $L_d^{(m)}>L$
\STATE \qquad \qquad \qquad Delete node $d \in \mathbb{S}_m$
\STATE \qquad \qquad \textbf{End if}
\STATE \qquad \textbf{End for}
\STATE \textbf{Until} there is no node in $\mathbb{S}_m$ or $m = M+1$
\STATE \textbf{Return} $M^* = m-1$ as the maximum throughput, and the path connecting node $1\in \mathbb{S}_0$ and node $d \in \mathbb{S}_{M^*}$ as the optimal solution
\end{algorithmic}
\end{algorithm}

\begin{example}\emph{
Suppose there are $3$ objects to be labelled, and two types of annotator clusters to be chosen from. The annotator and sub-channel budgets are given as $K = 6$ and $L = 3$, respectively. The first object can select type-$1$ cluster with $K_1 = 3$ and $L_1 = 1$, or type-$2$ cluster with $K_2 = 1$ and $L_1 = 3$, the corresponding resource uses are ($K_1^{(1)} = 3$, $L_1^{(1)} = 1$), and ($K_2^{(1)} = 1$, $L_2^{(1)} = 3$), respectively. The second object can select type-$1$ cluster with $K_1 = 3$ and $L_1 = 1$, or type-$2$ cluster with $K_2 = 1$ and $L_1 = 4$, the corresponding resource uses are ($K_1^{(2)} = 6$, $L_1^{(2)} = 2$), ($K_2^{(2)} = 4$, $L_2^{(2)} = 4$), ($K_3^{(2)} = 4$, $L_3^{(2)} = 5$), and ($K_4^{(2)} = 2$, $L_4^{(2)} = 7$), respectively. Since sub-channel use should be no larger than $3$, only node $1$ remains. The third object can select type-$1$ cluster with $K_1 = 3$ and $L_1 = 1$, or type-$2$ cluster with $K_2 = 1$ and $L_2 = 5$, the corresponding resource uses are ($K_1^{(3)} = 9$, $L_1^{(3)} = 3$), and ($K_2^{(3)} = 7$, $L_2^{(1)} = 7$), respectively. since annotator use should be no larger than $6$, there is no node in $\mathbb{S}_3$, thus the maximum throughput $M^* = 2$, and the optimal solution is the path connecting node-$1$ in $\mathbb{S}_0$ and node-$1$ in $\mathbb{S}_{2}$.
}
\end{example}

\begin{remark}[Complexity of Algorithm~\ref{Al:BBfading}]\emph{
Since there are $M$ levels and each level has at most $N^m$ edges, the complexity of Algorithm~\ref{Al:BBfading} is at most $\mathcal{O}(N^M)$, where $N$ and $M$ represent the number of cluster types and number of objects, respectively. However, the practical searching complexity should be less than $\mathcal{O}(N^M)$ due to the constraints on $K$ and $L$.
}
\end{remark}

\subsection{Special-Case Analysis}
To gain insights into the optimal design, we consider two special cases where either the annotator or the spectrum resource is constrained. For ease of exposition, the annotators are sorted in decreasing SNRs, i.e., $\gamma_1 \geq \gamma_2 \geq \dots \geq \gamma_K$. Then a throughput upper bound corresponds to the case where all the channel gains are as large as $g_1$, while the lower bound corresponds to the case that all the channel gains are as small as $g_K$. The corresponding sub-channel uses are
\begin{gather}
L_n^{\min} = \l\lceil \frac{\lambda_n S L}{\log_2 (1+\gamma_1) B T} \r\rceil, \label{minsubchannel} \\
L_n^{\max} = \l\lceil \frac{\lambda_n S L}{\log_2 (1+\gamma_K) B T} \r\rceil \label{maxsubchannel}.
\end{gather}

\subsubsection{Spectrum Constrained Case}
When the number of annotators is sufficiently large, the spectrum resource places the only constraint that limits the throughput. Combining Eq.~ \eqref{map}, \eqref{annotator} and \eqref{minsubchannel}, the criteria of spectrum constrained case is given in the following lemma.
\begin{lemma}[Criterion of Spectrum Constrained Case]\label{Lem:anno}
\emph{The optimization problem (P2) is spectrum constrained when
\begin{equation}\label{anno}
K \geq \max_{n} K_n \l\lfloor\frac{L}{L_n^{\min}} \r\rfloor,
\end{equation}
where $K_n$ and $L_n^{\min}$ are given in Eq.~\eqref{annotator} and \eqref{minsubchannel}, respectively. }
\end{lemma}
\proof
It can be observed from Eq.~\eqref{annotator} and \eqref{minsubchannel} that $K_n$ increases with the increasing $f(\lambda_n)$ in $[0,0.5]$, while $L_n^{\min}$ increases with the increasing $\lambda_n$. The criteria is achieved when the annotator budget can support the maximum annotator use.
\endproof

When the criteria of spectrum constrained case is met, the optimal design and the corresponding performance bounds are given in the proposition below.
\begin{proposition}[Optimal Design and Performance Bounds for Spectrum Constrained Case]\label{Prop:annofading}
\emph{The optimal design under the spectrum constrained condition should only assign the type-$N$ clusters to label all the objects, and the maximum throughput is bounded by
\begin{equation}
\l\lfloor\frac{L}{L_{N}^{\max}}\r\rfloor \leq M^* \leq \l\lfloor\frac{L}{L_{N}^{\min}}\r\rfloor.
\end{equation}
}
\end{proposition}
\proof
Due to the annotator-spectrum tradeoff, the type-$N$ clusters have the largest size and the smallest sub-channel use. Suppose that there exists another type of cluster denoted as $n'$, then the corresponding sub-channel use $L_{n'} \geq L_{N}$, thus the number of objects labelled by type-$n'$ cluster $M' = \l\lfloor\frac{L - L_N M^*}{L_{n'}}\r\rfloor \leq \l\lfloor\frac{L - L_N M^*}{L_N}\r\rfloor$. Since the annotators are sufficient, the type-$n'$ clusters should be replaced by the type-$N$ clusters. The lower and upper bounds are derived by replacing $L_N$ with $L_N^{\max}$ and $L_N^{\min}$, respectively.
\endproof

\subsubsection{Annotators Constrained Case}
When the number of sub-channels is sufficiently large, the annotator resource will become the only constraint that limits the throughput. Combining Eq.~\eqref{map}, \eqref{annotator} and \eqref{maxsubchannel}, the criteria of annotators constrained case is given in the following lemma.
\begin{lemma}[Criterion of Annotators Constrained Case]\label{Lem:carrier}
\emph{The optimization problem (P2) is annotators constrained when
\begin{equation}\label{carrier}
L \geq \max_{n} L_n^{\max} \l\lfloor\frac{K}{K_n} \r\rfloor,
\end{equation}
where $K_n$ and $L_n^{\max}$ are given in Eq.~\eqref{annotator} and \eqref{maxsubchannel}, respectively. }
\end{lemma}
 
The proof of Lemma~\ref{Lem:carrier} is similar to that of Lemma~\ref{Lem:anno} and thus omitted here. When the criteria of annotators constrained case is met, the optimal design should follow the proposition below.
\begin{proposition}[Optimal Design for Annotators Constrained Case]\label{Prop:carrier}
\emph{The optimal design under the annotators constrained condition should only assign the type-$1$ clusters to label all the objects. The corresponding maximum throughput $M^* = \l\lfloor\frac{K}{K_{1}}\r\rfloor$.}
\end{proposition}
\proof
Due to the annotator-spectrum tradeoff, the type-$1$ clusters have the smallest size and the largest sub-channel use. Suppose that there exists another type of cluster denoted as $n'$, then the corresponding annotator use $K_{n'} \geq K_1$, thus the number of objects labelled by type-$n'$ cluster $M' = \l\lfloor\frac{K - K_1 M^*}{K_{n'}}\r\rfloor \leq \l\lfloor\frac{K - K_1 M^*}{K_1}\r\rfloor$. Since the sub-channels are sufficient, the type-$n'$ clusters should be replaced by the type-$1$ clusters.
\endproof

\section{Optimal Design with Truncated Channel Inversion}
One reason for the complexity of the optimal JASA algorithm in Section IV is the heterogeneous gains of the annotator channels. As shown in this section, the complexity can be reduced if we use power control to equalize the channels by channel inversion. In such condition, finding the optimal solution can  be formulated as a knapsack problem that adopts an efficient solution method. Given the total power budget $P_t$ and sorting the annotators in the order of decreasing channel gains, i.e., $g_1 \geq g_2 \geq \dots \geq g_K$, the power allocation for truncated channel inversion is
\begin{equation}\label{Eq:power}
P_k = \begin{cases}
\frac{\gamma_0}{g_k},&k \leq K',\\
0,&k > K',
\end{cases}
\end{equation} 
where $\gamma_0$ represents the targeted SNR and the number of available annotators $K'$ is determined by power constraint $\sum_{k=1}^{K'} \frac{\gamma_0}{g_k} \leq P_t$, while other annotators are not deployed. The corresponding sub-channel uses are
\begin{equation}\label{carrierknap}
L_n = \l\lceil \frac{\lambda_n S L}{\log_2 (1+\gamma_0) B T} \r\rceil.
\end{equation}
One can see that the sub-channel uses are no longer differentiated by the cluster index $m$ as their counterparts without channel inversion. In other words, they are identical for the objects encoded in the same rate due to equalized channels. This leads to complexity reduction in searching for the optimal solution. Given Eq.~\eqref{carrierknap}, the problem (P2) can be reduced to 
\begin{equation*}\textbf{(P3)}\qquad
\begin{aligned}
\max_{\{X_n^{(m)}\}} \quad 
& \sum_{m=1}^M \mathbf{1} \Bigg\{\sum_{n=1}^N X_n^{(m)} > 0\Bigg\}\\ 
\text{s.t.} \quad 
& \sum_{m=1}^M \sum_{n=1}^N K_n X_n^{(m)} \leq K,\\
& \sum_{m=1}^M \sum_{n=1}^N L_n X_n^{(m)} \leq L,\\
& \sum_{n=1}^N X_n^{(m)} \in \{0,1\},~\forall~m,\\
& X_n^{(m)}  \in \{0,1\},~\forall~m,n.
\end{aligned}
\end{equation*}

\begin{remark}[Effect of Truncated Channel Inversion]\emph{On one hand, the exploitation of power optimization dimension may provide extra performance gain compared with the setting of uniform transmit power in Section IV. On the other hand, the truncated channel inversion will reduce the number of available annotators due to the SNR threshold and limited power resource. The effect of such tradeoff is further illustrated by simulation results and analyzed in Section VII.}
\end{remark}

\subsection{Knapsack Approach}
\begin{figure}[t]
\centering
\includegraphics[scale=0.55]{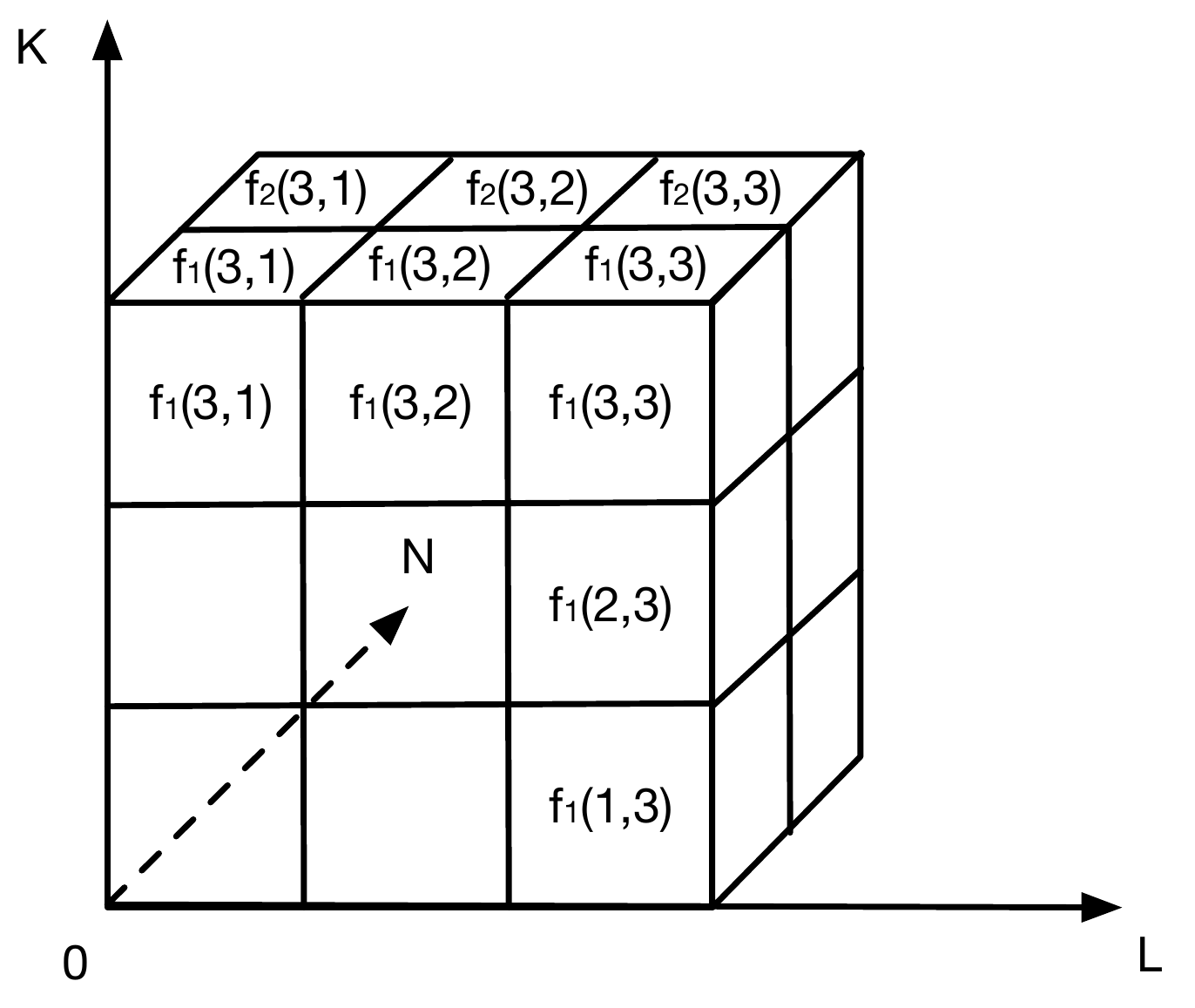}
\caption{Knapsack for throughput maximization with truncated channel inversion.}
\label{FigDPAWGN}
\end{figure}

Based on the truncated channel inversion design, an optimal solution approach can achieve pseudo-polynomial complexity by recasting the original problem (P3) into the following two-dimensional knapsack problem:
\begin{equation*}\textbf{(P4)}\qquad
\begin{aligned}
\max_{\{X_n\}} \quad 
& \sum_{n=1}^{N} X_n\\ 
\text{s.t.} \quad 
& \sum_{n=1}^{N} K_n X_n \leq K,\\
& \sum_{n=1}^{N} L_n X_n \leq L,\\
& 0 \leq X_n \leq \min \l\{\l\lfloor\frac{K}{K_n}\r\rfloor,\l\lfloor\frac{L}{L_n}\r\rfloor\r\},~\forall~n,
\end{aligned}
\end{equation*}
where $K_n$ and $L_n$ are given in \eqref{annotator} and \eqref{carrierknap}, respectively. The first two constraints give the budgets of annotators and sub-channels. The last constraint determines the maximum number of type-$n$ clusters that can be selected. 
According to \cite{kellerer2004multidimensional}, the knapsack problem (P4) is NP-complete but can be decomposed into a series of simpler sub-problems as demonstrated in Fig.~\ref{FigDPAWGN}, where $f_{\bar{n}}(k,u)$ represents the optimal solution of the $\bar{n}$-th sub-problem under annotator budget $k\in\{0,1,\dots,K\}$ and sub-channel budget $\ell\in\{0,1,\dots,L\}$. The first set of sub-problems is to maximize the throughput under the constraint that only type-$1$ cluster can be selected, i.e.,
\begin{equation*}\qquad
\begin{aligned}
\max_{X_1} \quad 
& X_1\\ 
\text{s.t.} \quad 
& L_1 X_1 \leq \ell,~ K_1 X_1 \leq k,\\
& 0 \leq X_1 \leq \min \l\{\l\lfloor\frac{k}{K_1}\r\rfloor,\l\lfloor\frac{\ell}{L_1}\r\rfloor\r\}.
\end{aligned}
\end{equation*}
The optimal solutions of the first set of sub-problems can be expressed as
\begin{equation}
f_1(k,\ell) = \min \l\{\l\lfloor\frac{k}{K_1}\r\rfloor,\l\lfloor\frac{\ell}{L_1}\r\rfloor\r\},
\end{equation}
where $k = 0,\dots,K$ and $\ell = 0,\dots,L$. The optimal solutions of $\bar{n}$-th set of sub-problems are determined based on the solutions of the previous sub-problems, expressed by
\begin{equation}
f_{\bar{n}}(k,\ell) = \max\l\{f_{\bar{n}-1}(k - m K_{\bar{n}}, \ell - m L_{\bar{n}}) + m|0 \leq m \leq \min \l\{\l\lfloor\frac{k}{K_{\bar{n}}}\r\rfloor,\l\lfloor\frac{\ell}{L_{\bar{n}}}\r\rfloor\r\}\r\},
\end{equation}
where $\bar{n} = 2,\dots,N$. If no type-$n$ cluster is selected, then $f_{\bar{n}}(k,\ell) = f_{\bar{n}-1}(k,\ell)$. If $m$ type-$n$ clusters are selected, then $f_{\bar{n}}(k,\ell) = f_{\bar{n}-1}(k - m K_{\bar{n}}, \ell - m L_{\bar{n}}) + m$ represents that part of old clusters are replaced by $m$ type-$\bar{n}$ clusters. To solve the sub-problems in a recursive manner, a low-complexity JASA algorithm via dynamic programming is shown in Algorithm~\ref{Al:DPAWGN}.

\begin{algorithm}[tt]
\renewcommand{\algorithmicrequire}{\textbf{Input:}}
\renewcommand{\algorithmicensure}{\textbf{Output:}}
\caption{Knapsack for Throughput Maximization with Truncated Channel Inversion.}
\label{Al:DPAWGN}
\begin{algorithmic}[1]
\REQUIRE annotator and sub-channel budget $K$ and $L$, annotator and sub-channel uses of type-$\bar{n}$ cluster $K_{\bar{n}}$ and $L_{\bar{n}}$
\ENSURE the maximum throughput $f_{\bar{n}}(k,u)$
\STATE \textbf{For} $\bar{n}=1:1:N$ \textbf{do}
\STATE \qquad \textbf{For} $k=0:1:K$ \textbf{do}
\STATE \qquad \qquad \textbf{For} $\ell=0:1:L$ \textbf{do}
\STATE \qquad \qquad \qquad  $f_{\bar{n}}(k,u) = \max\l\{f_{\bar{n}-1}(k - m K_{\bar{n}}, \ell - m L_{\bar{n}}) + m\r\}$
\STATE \qquad \qquad \textbf{End for}
\STATE \qquad \textbf{End for}
\STATE \textbf{End for}
\STATE \textbf{Return} $f_{\bar{n}}(k,u)$ as the maximum throughput
\end{algorithmic}
\end{algorithm}

\begin{remark}[Complexity of Algorithm~\ref{Al:DPAWGN}]\emph{
According to \cite{kellerer2004multidimensional}, the complexity of Algorithm~\ref{Al:DPAWGN} is pseudo-polynomial namely $\mathcal{O}(NKL)$,  where $N$, $K$, $L$ represent the number of cluster types, annotators, and sub-channels, respectively. }
\end{remark}

\begin{table*}[t]
\centering
\caption{Annotator and Sub-channel Uses at Each Node.}
\begin{tabular}{|c|c|c|c|c|c|c|c|c|}
\hline
$e \in \mathbb{S}_m$ & $1$ & $2$ & $\dots$ & $N$ & $N+1$ & $\dots$ & $C_{N+m-1}^m \!-\! 1$ & $C_{N+m-1}^m$ \\ 
\hline
$K_d^{(m)}$ & $m K_1$ & $(m\!-\!1) K_1 \!+\! K_2$ & $\dots$ & $(m\!-\!1) K_1 \!+\! K_N$ & $(m\!-\!2) K_1 \!+\! 2 K_2$ & $\dots$ & $K_{N\!-\!1} \!+\! (m\!-\!1) K_N$ & $ m K_N$ \\ 
\hline
$L_d^{(m)}$ & $m L_1$ & $(m\!-\!1) L_1 \!+\! L_2$ & $\dots$ & $(m\!-\!1) L_1 \!+\! L_N$ & $(m\!-\!2) L_1 \!+\! 2 L_2$ & $\dots$ & $L_{N\!-\!1} \!+\! (m\!-\!1) L_N$ & $ m L_N$ \\ 
\hline
\end{tabular}
\label{order2}
\end{table*}

\begin{figure}[t]
\centering
\includegraphics[scale=0.55]{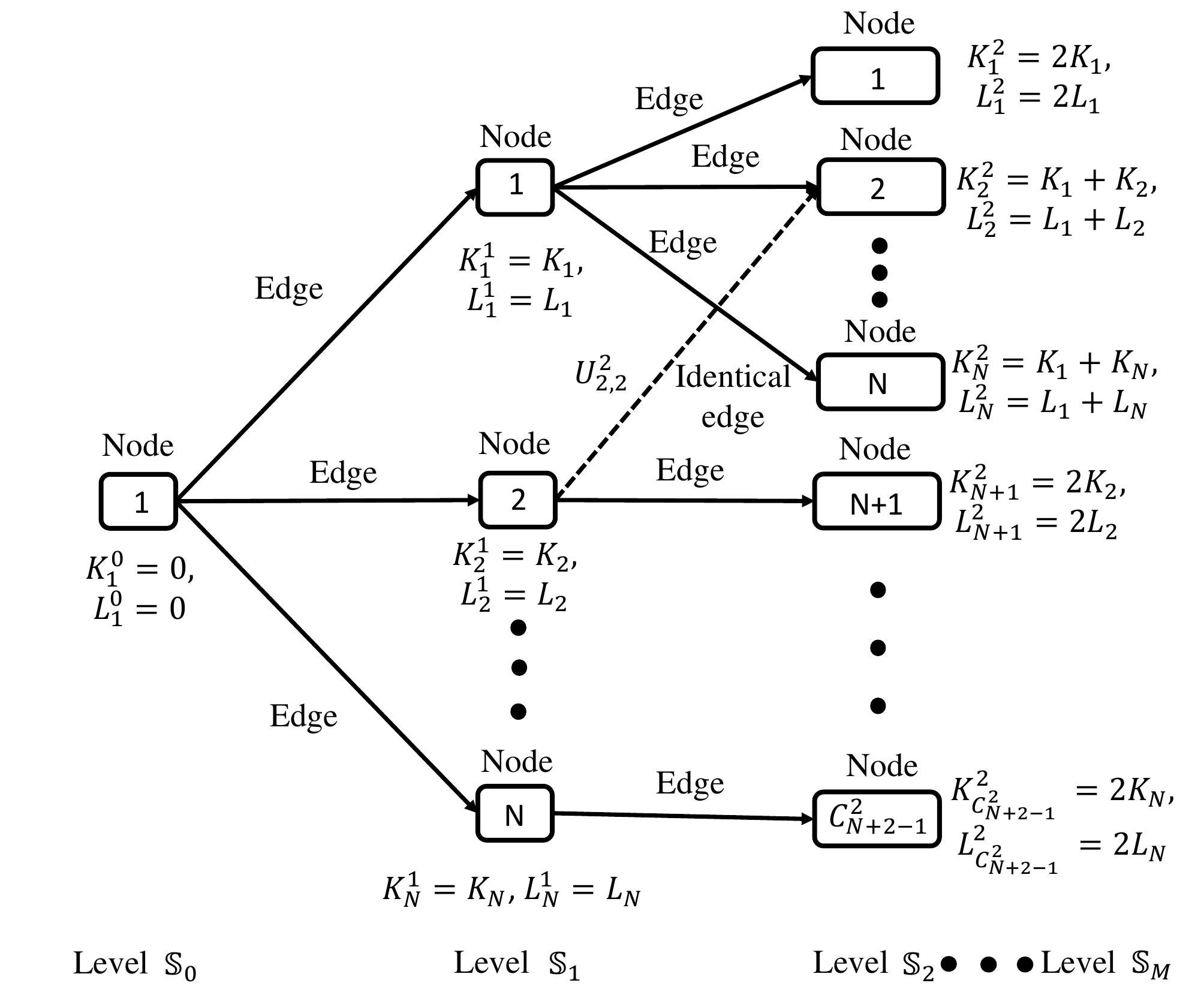}
\caption{Tree search for throughput maximization with truncated channel inversion.}
\label{FigBBAWGN}
\end{figure}

\subsection{Tree Search Approach}
An alternative approach for solving the knapsack problem is via tree search \cite{frieze1976shortest}. Here the tree search approach discussed in Section IV can be simplified with truncated channel inversion, and the resultant homogeneous resource allocation to objects. Specifically, some nodes in Fig.~\ref{FigBBfading} are identical (e.g., the node $2$ and node $N+1$ in level $\mathbb{S}_2$) and thus can be merged as shown in Fig.~\ref{FigBBAWGN}. The reduced number of nodes in level $\mathbb{S}_m$ is given below.
\begin{lemma}[Graph Truncation]\label{Lem:Merge}
\emph{In level $\mathbb{S}_m$, the number of reduced nodes due to merging is $N^m - C_{N+m-1}^m$.
}
\end{lemma}
\proof
Since there are $M$ objects to be labelled and each object can be labelled by $N$ types of clusters, the number of nodes is $N^M$ if considering the clustering sequence. Under the equivalent SNRs, the clustering sequence is no longer considered and thus the number of nodes only depends on the possible combinations with repetitions, which is $C_{N+M-1}^M$ according to \cite{benjamin2003proofs}. 
\endproof
 
Moreover, for those edges having the identical rewards and leading to the same nodes (e.g., $U_{1,1}^1 + U_{1,2}^2 = U_{1,2}^1 + U_{2,2}^2$ and both two edges lead to node $2\in S_2$) are defined as the \emph{identical edges}. If two edges are identical then one of the edges can be deleted to improve the search efficiency. Without loss of generality, only the first one among the identical edges is kept and the nodes are listed in Table~\ref{order2}. To derive the optimal solution, a low-complexity JASA algorithm via branch-and-bound is shown in Algorithm~\ref{Al:BBAWGN}.

\begin{algorithm}[tt]
\renewcommand{\algorithmicrequire}{\textbf{Input:}}
\renewcommand{\algorithmicensure}{\textbf{Output:}}
\caption{Tree Search for Throughput Maximization with Truncated Channel Inversion.}
\label{Al:BBAWGN}
\begin{algorithmic}[1]
\REQUIRE number of objects to be labelled $M$, annotator and sub-channel budgets $K$ and $L$
\ENSURE the maximum throughput $M^*$, and the optimal solution
\STATE Initialize $m=0$
\STATE \textbf{Loop} $m = m+1$
\STATE \qquad \textbf{For} $d = 1:1:|\mathbb{S}_m|$
\STATE \qquad \qquad Create the $d$-th node with $K_d^{(m)}$ and $L_d^{(m)}$ according to Table~\ref{order2}
\STATE \qquad \qquad \textbf{If} $K_d^{(m)}>K$ or $L_d^{(m)}>L$
\STATE \qquad \qquad \qquad Delete node $d \in \mathbb{S}_m$
\STATE \qquad \qquad \textbf{End if}
\STATE \qquad \textbf{End for}
\STATE \textbf{Until} there is no node in $\mathbb{S}_m$ or $m = M+1$
\STATE \textbf{Return} $M^* = m-1$ as the maximum throughput, and the path connecting node $1\in \mathbb{S}_0$ and node $d \in \mathbb{S}_{M^*}$ as the optimal solution
\end{algorithmic}
\end{algorithm}

\begin{remark}[Complexity Comparison between Knapsack and Tree Search Approaches]\emph{
Since there are M levels and at most $C_{N+m-1}^m$ nodes in node $S_m$, the complexity of Algorithm~\ref{Al:BBAWGN} is at most $\mathcal{O}\l(C_{N+M-1}^M\r) \approx \mathcal{O}\l(\l(\frac{N+M-1}{M}e\r)^M\r)$, where $N$ and $M$ represent the number of cluster types and the number of objects, respectively. However, the practical searching complexity should be less due to the constraints on $K$ and $L$. It can be observed that the tree search approach is more efficient when the number of objects is small enough that satisfying $\l(\frac{N+M-1}{M}e\r)^M \leq NKL$, and vice versa.
}
\end{remark}

\subsection{Special-Case Analysis}
Again, we study the optimal policy in the current case for the spectrum/annotators constrained cases as in Section III. 

\subsubsection{Spectrum Constrained Case} When the spectrum resource is constrained, the problem (P4) can be simplified as

\begin{equation*}\textbf{(P5)}\qquad
\begin{aligned}
\max_{\{X_n\}} \quad 
& \sum_{n=1}^{N} X_n\\ 
\text{s.t.} \quad 
& \sum_{n=1}^{N} L_n X_n \leq L,\\
& 0 \leq X_n \leq \l\lfloor\frac{L}{L_n}\r\rfloor,~\forall~n.
\end{aligned}
\end{equation*}

Through the similar derivation of Lemma~\ref{Lem:anno}, the criteria of spectrum constrained case and the corresponding optimal design are given in the following lemma.
\begin{lemma}[Criterion and Optimal Solution in Spectrum Constrained Case]\label{Lem:annotatorTCI}
\emph{The optimization problem is spectrum constrained when
\begin{equation}
K \geq \max_{n} K_n \l\lfloor\frac{L}{L_n}\r\rfloor,
\end{equation}
where $K_n$ and $L_n$ are given in Eq.~\eqref{annotator} and \eqref{carrierknap}, respectively. The optimal design under spectrum constrained condition should only contain the type-$N$ clusters for all objects. The corresponding maximum throughput $M^* = \l\lfloor\frac{L}{L_{N}}\r\rfloor$.}
\end{lemma}

\subsubsection{Annotators Constrained Case} When the annotator resource is constrained, the problem (P4) can be simplified as
\begin{equation*}\textbf{(P6)}\qquad
\begin{aligned}
\max_{\{X_n\}} \quad 
& \sum_{n=1}^{N} X_n\\ 
\text{s.t.} \quad 
& \sum_{n=1}^{N} K_n X_n \leq K,\\
& 0 \leq X_n \leq \l\lfloor\frac{K}{K_n}\r\rfloor,~\forall~n.
\end{aligned}
\end{equation*}

Through the similar derivation of Lemma~\ref{Lem:carrier}, the criteria of annotators constrained case and the corresponding optimal solution under fading channels are given in the following lemma.
\begin{lemma}[Criterion and Optimal Solution in Annotators Constrained Case]\label{Lem:carrierTCI}
\emph{The optimization problem is annotators constrained when
\begin{equation}
L \geq \max_{n} L_n \l\lfloor\frac{K}{K_n} \r\rfloor,
\end{equation}
where $K_n$ and $L_n$ are given in Eq.~\eqref{annotator} and \eqref{carrierknap}, respectively. The corresponding optimal clustering strategy should only contain the type-$1$ clusters for all objects, and the maximum throughput can be expressed as $M^* = \l\lfloor\frac{K}{K_1}\r\rfloor$.
}
\end{lemma}

\section{Extension to Frequency Selective Fading}
The preceding designs are based on the assumption of frequency non-selective channels. Consider the scenario of frequency selective channels where channel gains vary over sub-channels. The variation gives rise to a matching problem between sub-channels and objects. The size of data that can be received by the $k$-th annotator in cluster $\mathbb{K}_m$ can be expressed by
\begin{equation}
Z_{k}  = \sum_{\ell=1}^{L} \frac{\rho_{m,\ell} B}{L} \log_2 (1+\frac{|g_{k,\ell}|^2 P_k}{N_0}),
\end{equation}
where $g_{k,\ell}$ represents the channel gain in the $\ell$-th sub-channel from the AP to the $k$-th annotator, and $\rho_{m,\ell} \in \{0,1\}$ indicates whether the $\ell$-th sub-channel is allocated for multicasting the object $\hat{o}_m$. For the previous scenario of frequency non-selective fading, the sub-channel allocation for multicasting an object is based on the worst channel gain in the cluster. In the current scenario, this is no longer valid since each channel now comprises multiple gains for its sub-channels. As a result, the optimal sub-channel allocation is changed from merely determining the number of sub-channels in the previous scenario to solving a complex problem of matching each object to a subset of sub-channels in the current scenario. A simple but sub-optimal approach for extending the designs in the preceding sections is as follows. One can sort the annotators according to the average effective channel power gains for clustering as indicated by Proposition~\ref{Optimal}, and then solve the matching problem by applying the game-theoretic based algorithms \cite{roth1992two}. The extension is straight forward and the detailed design is out of the scope of this paper.

\section{Simulation Results}
In this section, the performance of the wireless crowd labelling framework is evaluated by simulations. The simulation parameters are set as follows unless specified otherwise. There are $N = 3$ types of encoding rates denoted as $\lambda_n \in \{0.5\log_2 3, 0.5, 0.5\log_2 1.5\}$, with the corresponding LEPs $f(\lambda_n) \in \{0.1, 0.15, 0.2\}$. The RLEP threshold is set as $Q = 0.1$. Then the required annotator cluster sizes are $K_n \in \{1, 3, 5\}$, corresponding to the setting of LEP. The data-source variance is set as $\sigma^2 = 0.3$ and thus the encoding rates. All the channels are assumed to be i.i.d. \emph{Rayleigh fading}. Without loss of generality, the symbols of each object, unit bandwidth, channel noise, and latency are set as $S = 10$, $B/L = 3$, $N_0 = 1$, and $T = 1$, respectively. The total transmit power budget at the AP is assumed to equal to the number of annotators. For truncated channel inversion, the targeted SNR is set as $\gamma_0 = 1$.

\begin{figure}[t]
  \centering
  \subfigure[Throughput versus number of annotators.]{
  \label{FigFadingK}
  \includegraphics[scale=0.55]{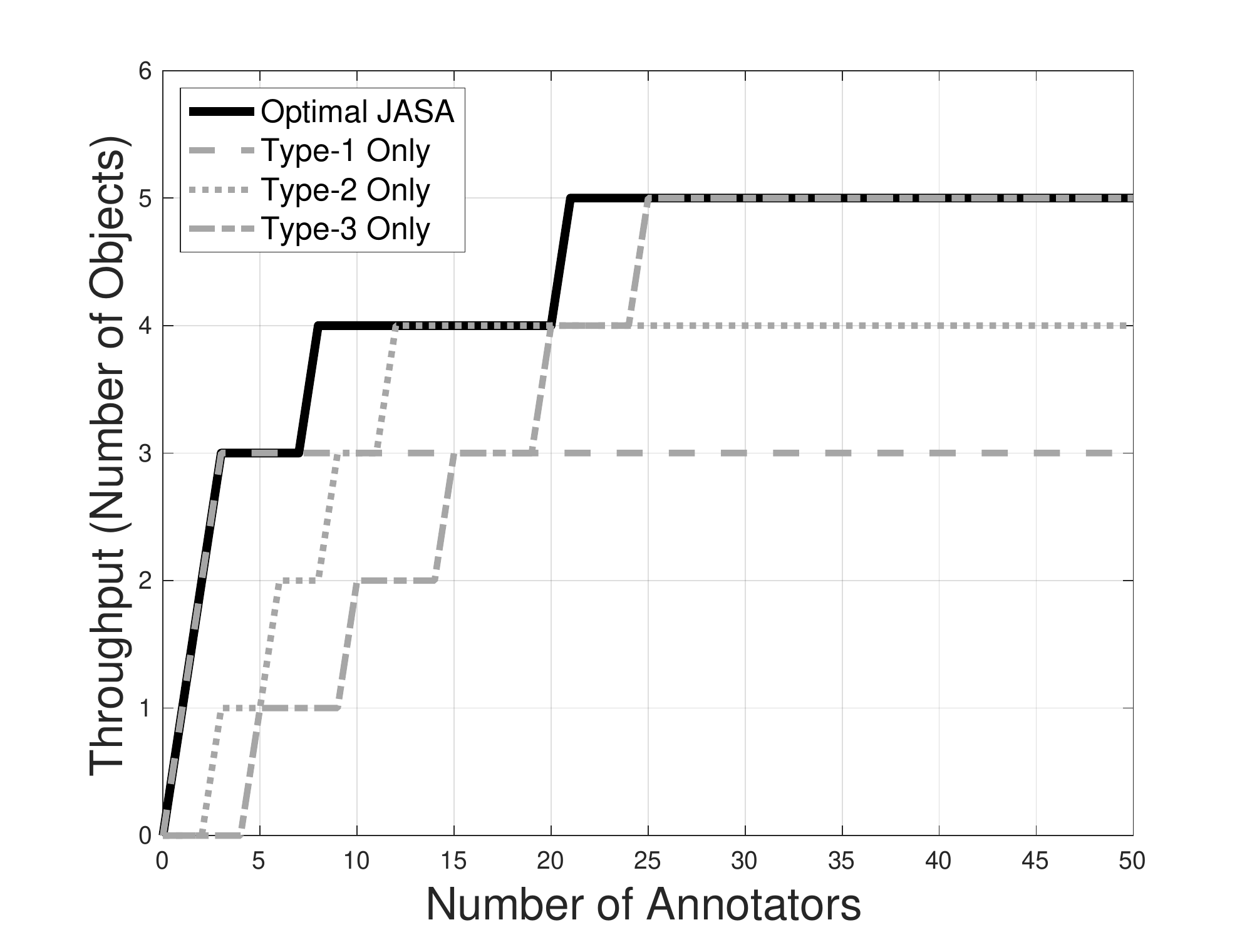}}
  \subfigure[Throughput versus number of sub-channels.]{
  \label{FigFadingL}
  \includegraphics[scale=0.55]{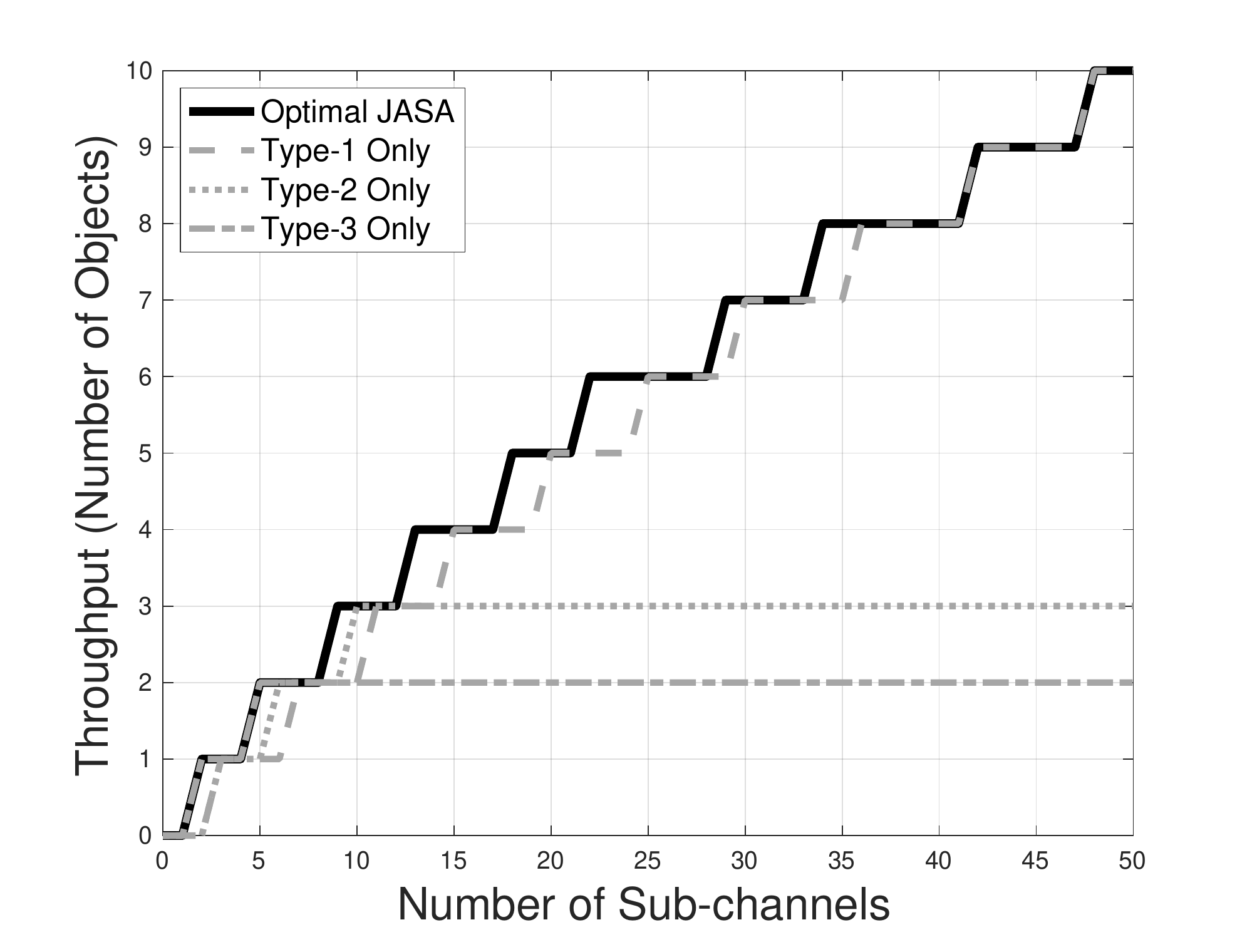}}
  \caption{Throughput maximization in fading channel case.}
  \label{FigFading}
\end{figure}

\subsection{Throughput Maximization in Fading Channel Case}
In fading channel case, four designs are considered for our performance comparison. The first one is the \emph{optimal JASA} algorithm based on branch-and-bound, which applies the tree search to find the optimum with exponential complexity. The other three benchmark designs only use one of $N$ types of clusters, named \emph{type-1/2/3 only} designs, respectively. First, the curves for the throughput versus the number of annotators are displayed in Fig.~\ref{FigFadingK} with the sub-channel budget $L = 10$. It can be observed that the throughput achieved by the \emph{optimal JASA} algorithm first increases with the increasing number of annotators. However, when the annotator budget exceeds a threshold (of about 21), the performance cannot improve further by increasing the annotator budget. The reason is that the bottleneck in this case is no longer annotators but other settings, such as the sub-channel budget. Moreover, the \emph{optimal JASA} algorithm can always achieve the optimum, while other three designs can achieve the optimums in particular phases. Specifically, the \emph{type-1 only} design is optimal under the shortage of annotators ($K \leq 7$), while the \emph{type-3 only} design is optimal when the annotators are sufficient ($K \geq 25$), and the \emph{type-2 only} design is optimal when the number of annotators and sub-channels are comparative.

Next, the curves for the throughput versus the number of sub-channels are displayed in Fig.~\ref{FigFadingL} with the annotator budget $K = 10$. It can be observed that the throughput achieved by the \emph{optimal JASA} algorithm first increases with the increasing sub-channels and then tends to converge after $L = 48$, since the constrained resource is no longer sub-channels. The optimality analysis is similar to the one for Fig.~\ref{FigFadingK}. Specifically, the \emph{type-3} only design is optimal under the shortage of sub-channels ($L \leq 8$), while the \emph{type-1 only} design is optimal when the sub-channels are sufficient ($L \geq 36$), and the \emph{type-2 only} design is optimal when the number of annotators and sub-channels are comparative.

\begin{figure}[t]
  \centering
  \subfigure[Throughput versus number of annotators.]{
  \label{FigAWGNK}
  \includegraphics[scale=0.55]{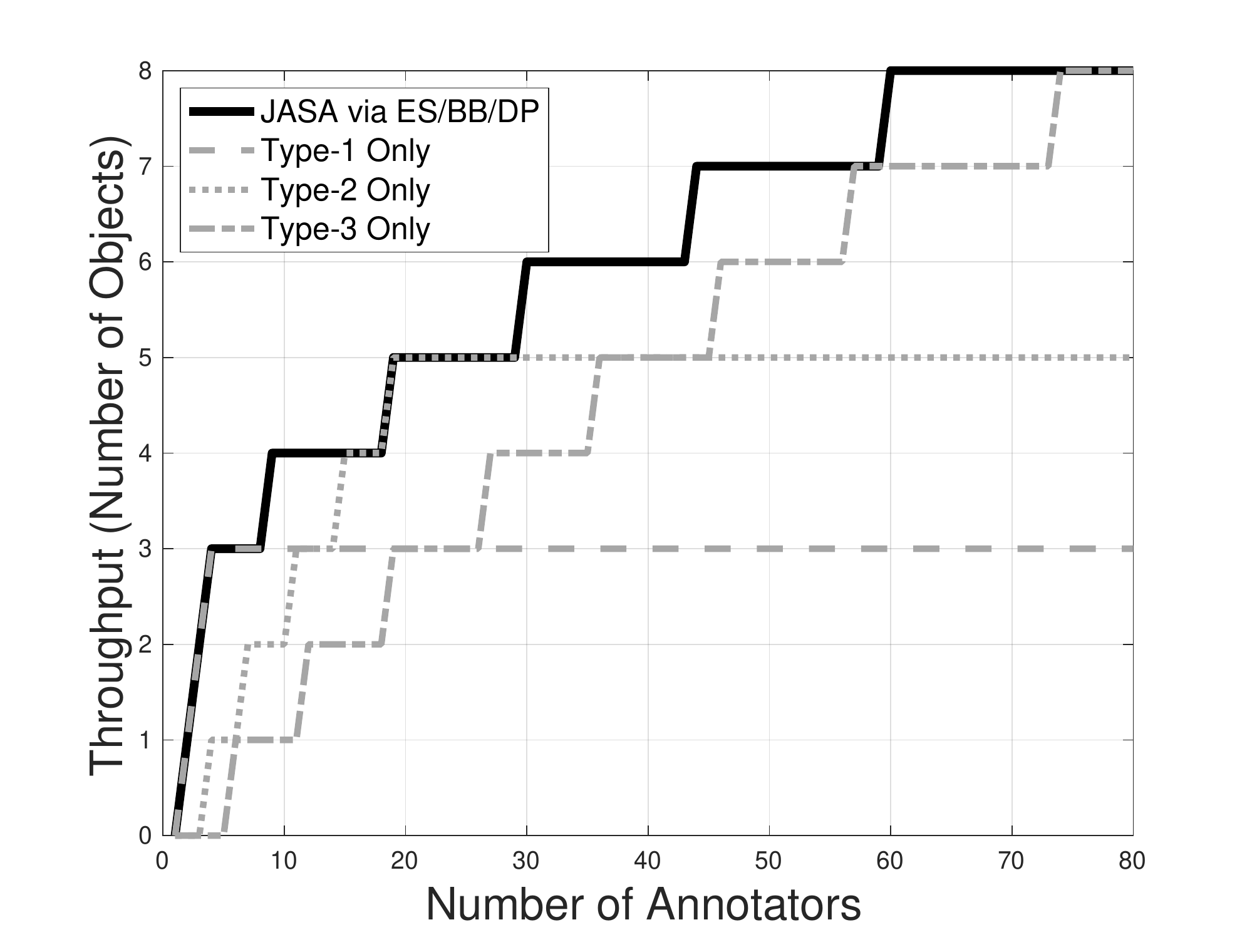}}
  \subfigure[Throughput versus number of sub-channels.]{
  \label{FigAWGNL}
  \includegraphics[scale=0.55]{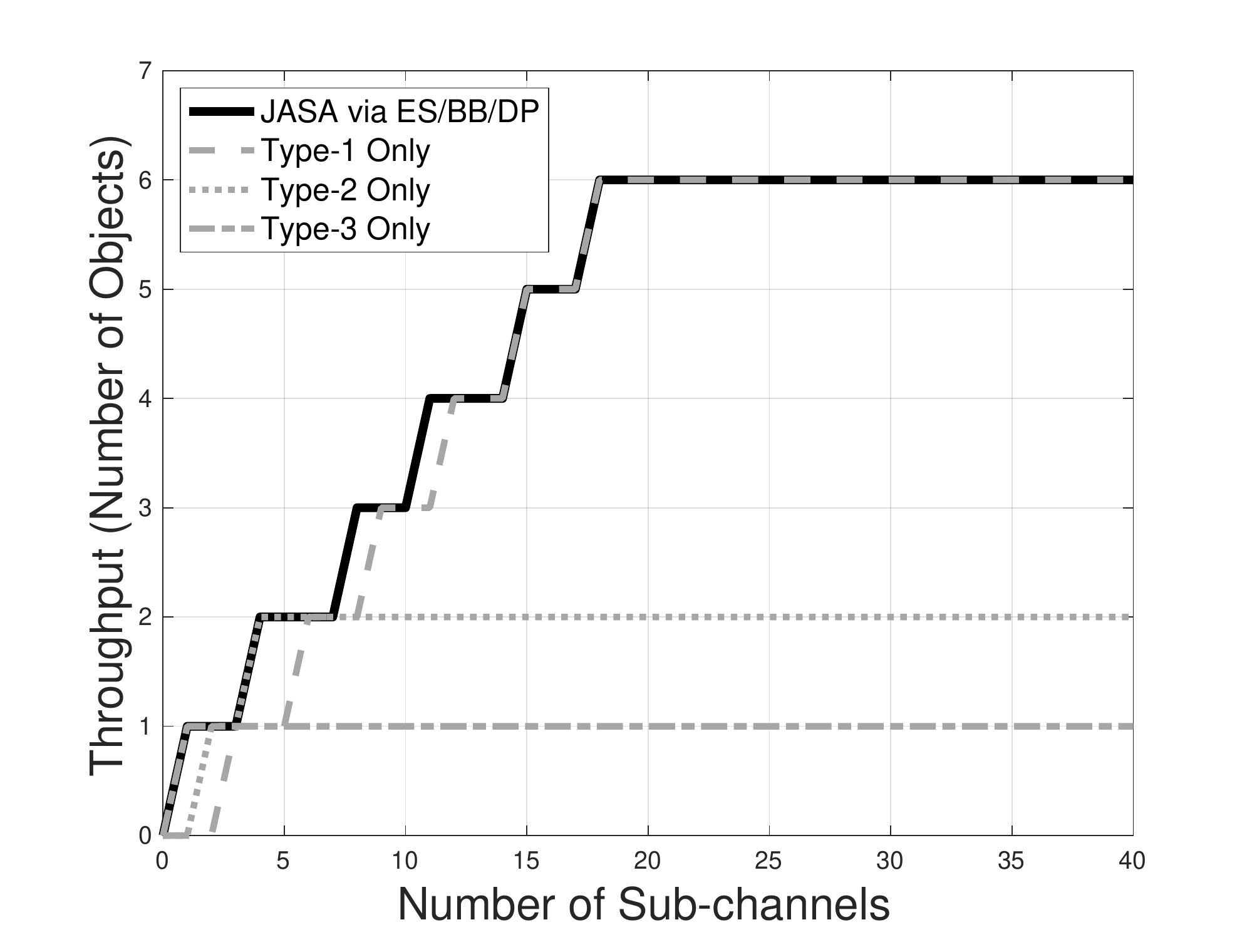}}
  \caption{Throughput maximization with truncated channel inversion.}
  \label{FigAWGN}
\end{figure}

\subsection{Throughput Maximization based on Truncated Channel Inversion}
After truncated channel inversion, six designs are considered for the performance comparison. The low-complexity \emph{JASA algorithm via BB} applies branch-and-bound search on the merged and truncated path graph, while the low-complexity \emph{JASA algorithm via DP} solves a series of sub-problems recursively based on dynamic programming, and the \emph{JASA algorithm via ES} simply traverses all feasible solutions through exhausted search. The curves for the throughput versus the number of annotators are displayed in Fig.~\ref{FigAWGNK} with the sub-channel budget $L = 10$. It can be observed that the low-complexity \emph{JASA algorithms via BB and DP} can achieve the same performance as the one via ES, which is the optimum for the truncated channel inversion case. The \emph{type-1 only} design can achieve the optimum under the annotator constrained case ($K \leq 8$), and the \emph{type-3 only} design can achieve the optimum when the annotators are sufficient ($K \geq 74$). Moreover, by comparing Fig.~\ref{FigFadingK} with Fig.~\ref{FigAWGNK}, it can be observed that the throughput after truncated channel inversion is larger than the one with fixed power when $K \geq 30$, which indicates that the performance gain introduced by the dimension of power optimization is larger than the performance loss due to the reduced number of annotators when the annotators are sufficient.

Fig.~\ref{FigAWGNL} further demonstrates the curves for the throughput versus the number of sub-channels with the annotator budget $K = 10$. It can be observed that the three \emph{JASA algorithms via BB, DP, and ES} can all achieve the optimum of the truncated channel inversion case, while the \emph{type-3 only} design can achieve the optimum under the sub-channel constrained case ($L \leq 3$), and the \emph{type-1 only} design can achieve the optimum when the sub-channels are sufficient ($L \geq 12$). Moreover, by comparing Fig.~\ref{FigFadingL} with Fig.~\ref{FigAWGNL}, it can be observed that the throughput after truncated channel inversion is smaller than the one with fixed power when $L \geq 22$, which indicates that the performance gain introduced by the dimension of power optimization cannot compensate for the loss due to the reduced number of annotators when the annotators are constrained.

\begin{figure}[t]
  \centering
  \subfigure[Running time versus number of annotators.]{
  \label{FigtimeK}
  \includegraphics[scale=0.55]{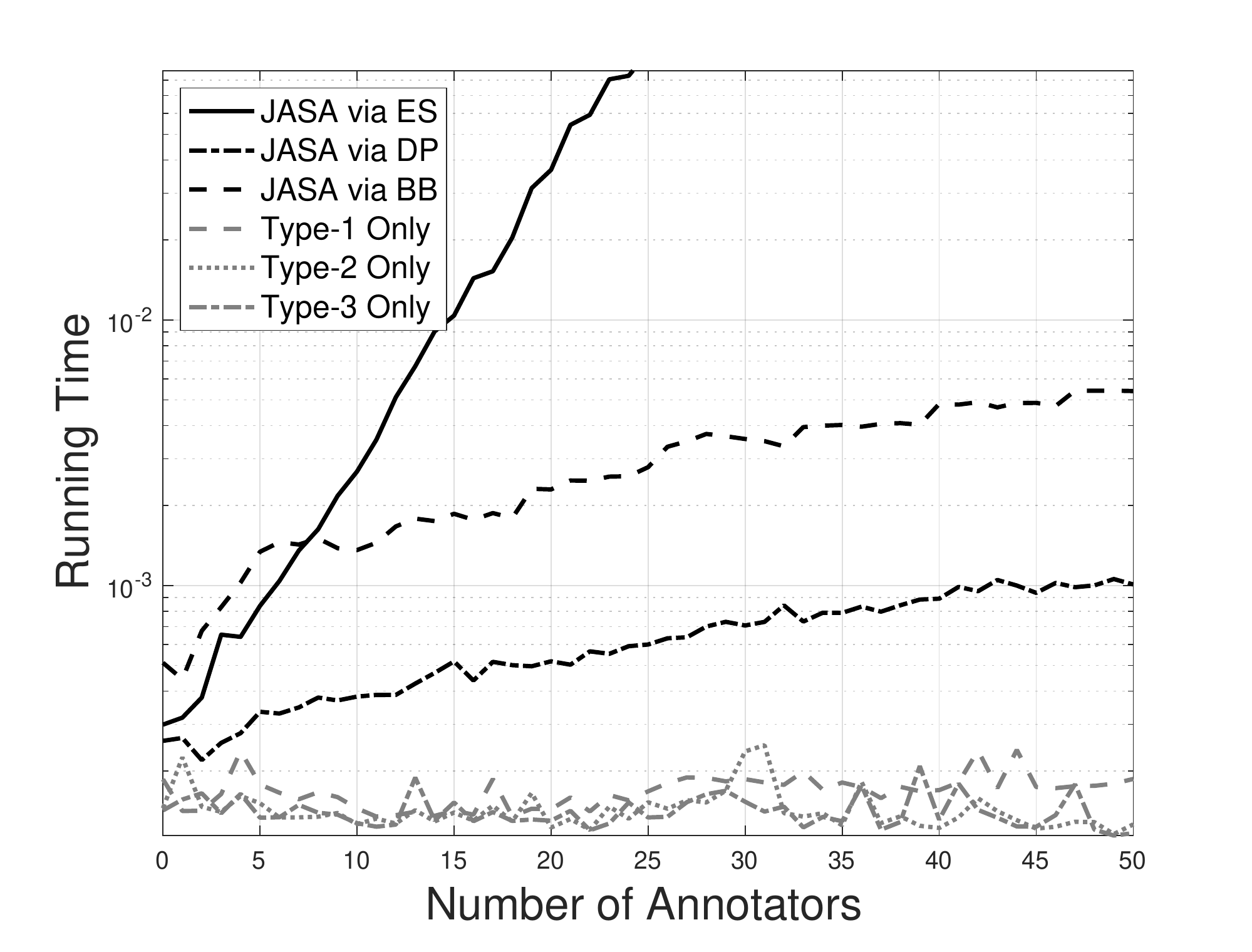}}
  \subfigure[Running time versus number of sub-channels.]{
  \label{FigtimeL}
  \includegraphics[scale=0.55]{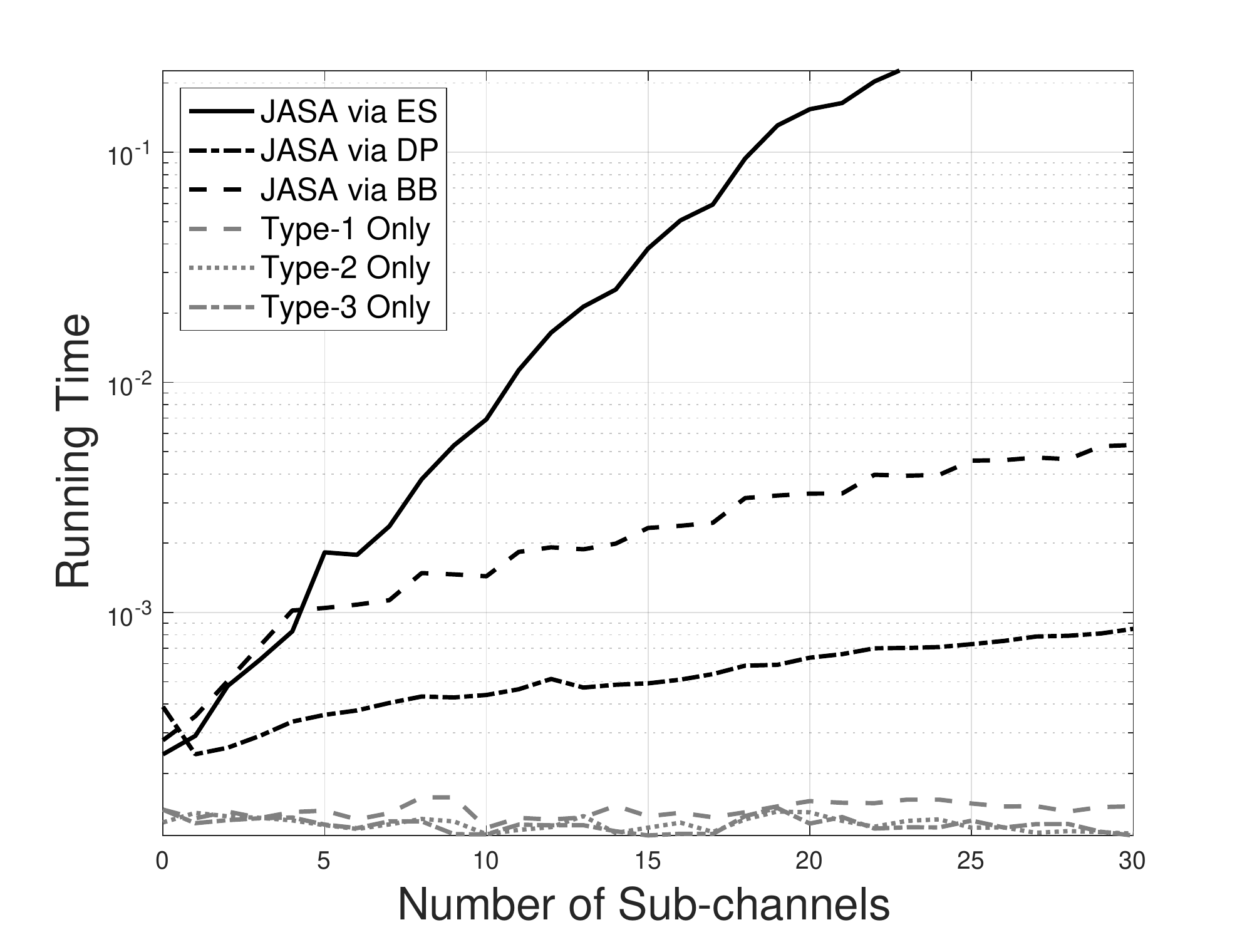}}
  \caption{Running time based on TCI.}
  \label{Figtime}
\end{figure}

\subsection{Running Time based on Truncated Channel Inversion}
To compare the complexities of our proposed algorithms, the practical running times are tested and recorded. Fig.~\ref{FigtimeK} illustrates the relationship between the running time and the number of annotators with the sub-channel budget $L = 15$. It can be observed that the running times of both the \emph{JASA algorithms via ES and BB} increase exponentially with the increasing number of annotators, while the increasing trends of the one via BB is relatively slower. The running time of the \emph{JASA algorithm via DP} increases linearly with the increasing number of annotators, which is much slower than the algorithms via ES and BB. Under the situation with the limited number of annotators, the running time of the \emph{JASA algorithm via DP} is even comparable with that of other three simple designs, which simply use one of $N$ types of clusters.

The relationship between running time and the number of sub-channels is further illustrated in Fig.~\ref{FigtimeL} with the annotator budget $K = 20$. It can be observed that the running times of the three \emph{JASA algorithms via BB, DP, and ES} all increases with the number of sub-channels, while the other observations are similar to those in Fig.~\ref{FigtimeK}.

\subsection{Effect of Targeted SNR}
\begin{figure}[t]
\centering
\includegraphics[scale=0.55]{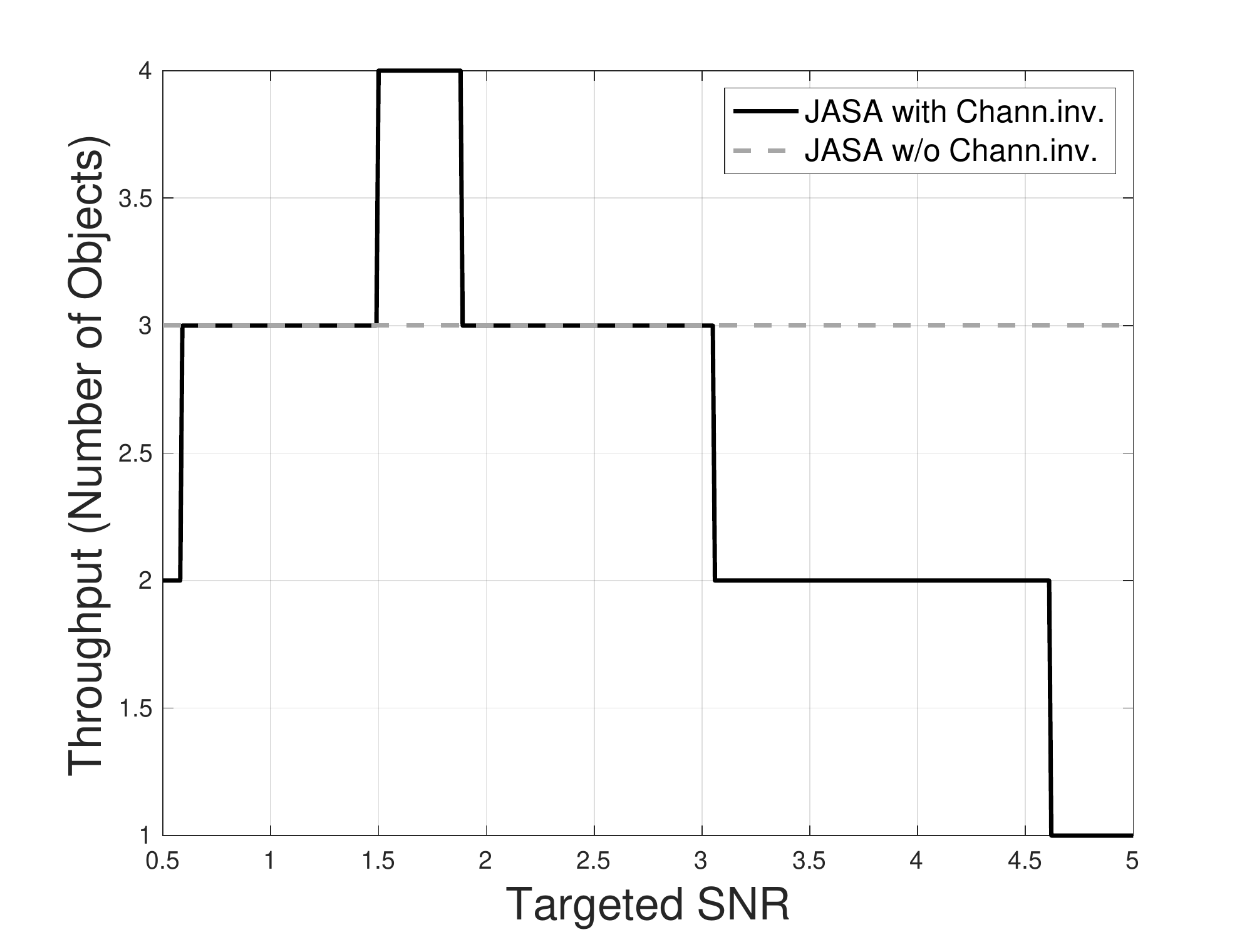}
\caption{Throughput versus targeted SNR.}
\label{FigSNR}
\end{figure}

To illustrate the effect of targeted SNR on the performance, the curves for the throughput versus the target SNR are displayed in Fig.~\ref{FigSNR} with the annotator budget $K = 10$ and the sub-channel budget $L = 10$. The optimal throughput under fading channels and after truncated channel inversion are presented by the curves \emph{JASA w/o Chann.inv} and \emph{JASA with Chann.inv}, respectively. It can be observed that the throughput after truncated channel inversion is first increasing and then decreasing with the growth of the targeted SNR, while the throughput under the fading channels is irrelevant with the targeted SNR and thus remains as a constant. Specifically, when the targeted SNR is extremely small ($0.5 \leq \gamma_0 \leq 0.6$), it is easy to be achieved by all the annotators but will cause severe sub-channels consumption for the data transmission, thus the performance of \emph{JASA with Chann.inv} is not as good as \emph{JASA w/o Chann.inv}. When the targeted SNR is relatively small ($1.5 \leq \gamma_0 \leq 1.9$), it is still easy to be achieved by most of the annotators and can save the consumption of sub-channels comparing with the one without truncated channel inversion, thus the performance of \emph{JASA with Chann.inv} is better. When the targeted SNR is relatively large ($\gamma_0 \geq 3.1$), more and more annotators cannot achieve such value and become unavailable, finally there is no available annotators due to the high targeted SNR and thus no object can be labelled.

\vspace{-5mm}
\section{Concluding Remarks}
This work proposed a wireless crowd labelling framework and investigated a joint design of encoding rate, annotator clustering, and sub-channel allocation. The framework design is tractable by constructing a search tree of the original NP-hard combinatorial problem and searching the optimal solution via the algorithm based on branch-and-bound. Two more efficient algorithm designs based on branch-and-bound and dynamic programming are further introduced to derive the optimal solution after truncated channel inversion. The performances of the proposed algorithms are further evaluated by the simulation results. This work points to a promising new research area of wireless crowd labelling where many interesting research issues warrant further investigation, such as power allocation, spatial beamforming, and annotator-channel matching.

\section*{Acknowledgment}
The work of K. Huang is supported by the Hong Kong Research Grants Council under the Grants 17208319 and 17209917. The work of Y. Gong is supported by the Shenzhen Science and Technology Program under Grants JCYJ20170817110410346, JSGG20180508151852303, and the Peng Cheng Laboratory under Grant PCL2018KP002. The work of K. Shen and W. Yu is supported by the Natural Sciences and Engineering Research Council (NSERC) of Canada via the Canada Research Chairs Program.

\appendix
\subsection{Proof of Proposition \ref{Optimal}}\label{App:Optimal}
Suppose that there are $K$ annotators arranged in the order of decreasing channel gains for labelling two objects. Suppose there are two encoding rates satisfying $\lambda_a \geq \lambda_b$, here two cases are considered. In the first case, suppose the first object is encoded at rate $\lambda_a$ and the second one is $\lambda_b$, thus the first $K_a$ annotators are clustered as $\mathbb{K}_1$, and the remaining $K_b = K - K_a$ annotators are clustered as $\mathbb{K}_2$. Based on Eq.~\eqref{size} and \eqref{reception}, the total sub-channel use is
\begin{equation*}
L_{\text{case 1}}^* = \left\lceil\frac{S \lambda_a L}{\log_2 (1+g_{K_a}P/N_0) BT}\right\rceil + \left\lceil\frac{S \lambda_b L}{\log_2 (1+g_{K}P/N_0) BT}\right\rceil.
\end{equation*} 
If any annotator in cluster $\mathbb{K}_1$ exchanges cluster with one in $\mathbb{K}_2$ denoted as $k'$ ($k' \neq K$), the corresponding sub-channel use is 
\begin{equation*}
L_{\text{case 1}}' = \left\lceil\frac{S \lambda_a L}{\log_2 (1+g_{k'}P/N_0) BT}\right\rceil + \left\lceil\frac{S \lambda_b L}{\log_2 (1+g_{K}P/N_0) BT}\right\rceil.
\end{equation*} 
Since $g_{K_a} \geq g_{k'}$, we have $L_{\text{case 1}}^* \leq L_{\text{case 1}}'$. If any annotator in cluster $\mathbb{K}_1$ exchanges cluster with the $K$-th annotator, the corresponding sub-channel use is 
\begin{equation*}
L_{\text{case 1}}'' = \left\lceil\frac{S \lambda_a L}{\log_2 (1+g_{K}P/N_0) BT}\right\rceil + \left\lceil\frac{S \lambda_b L}{\log_2 (1+g_{K-1}P/N_0) BT}\right\rceil.
\end{equation*} 
Without the specified $\lambda_a$ and $\lambda_b$, it is hard to compare $L_{\text{case 1}}''$ with $L_{\text{case 1}}^*$. However, consider the second case where the first object is encoded at rate $\lambda_b$ and the second one is $\lambda_a$, thus the first $K_b$ annotators are clustered as $\mathbb{K}_1$, and the remaining $K_a = K - K_b$ annotators are clustered as $\mathbb{K}_2$. The corresponding sub-channel use is
\begin{equation*}
L_{\text{case 2}}^* = \left\lceil\frac{S \lambda_b L}{\log_2 (1+g_{K_b}P/N_0) BT}\right\rceil + \left\lceil\frac{S \lambda_a L}{\log_2 (1+g_{K}P/N_0) BT}\right\rceil.
\end{equation*} 
Since $g_{K_b} \geq \gamma_{K-1}$, we have $L_{\text{case 2}}^* \leq L_{\text{case 1}}''$. Similarly, if any annotator in cluster $\mathbb{K}_1$ exchanges cluster with one in $\mathbb{K}_2$ denoted as $k'$ ($k' \neq K$), the corresponding sub-channel use is 
\begin{equation*}
L_{\text{case 2}}' = \left\lceil\frac{S \lambda_b L}{\log_2 (1+g_{k'}P/N_0) BT}\right\rceil + \left\lceil\frac{S \lambda_a L}{\log_2 (1+g_{K}P/N_0) BT}\right\rceil.
\end{equation*} 
Since $g_{K_b} \geq g_{k'}$, we have $L_{\text{case 2}}^* \leq L_{\text{case 2}}'$. If any annotator in cluster $\mathbb{K}_1$ exchanges cluster with the $K$-th annotator, the corresponding sub-channel use is 
\begin{equation*}
L_{\text{case 2}}'' = \left\lceil\frac{S \lambda_b L}{\log_2 (1+g_{K}P/N_0) BT}\right\rceil + \left\lceil\frac{S \lambda_a L}{\log_2 (1+g_{K-1}P/N_0) BT}\right\rceil.
\end{equation*} 
Since $g_{K_a} \geq g_{K-1}$, we have $L_{\text{case 1}}^* \leq L_{\text{case 2}}''$. Given identical total annotator use, the strategy with the least sub-channel use leads to the maximum throughput. Thus the optimal solution should be either $L_{\text{case 1}}^*$ or $L_{\text{case 2}}^*$, both of which are resulted from sequentially annotators clustering in the order of decreasing channel gains.

\bibliographystyle{ieeetr}

\end{document}

%% file: header.tex
\newtheorem{lemma}{Lemma}

\newtheorem{proposition}{Proposition}

\newtheorem{example}{\bf Example}
\newtheorem{remark}{\bf Remark}

\def\qed{$\Box$}

\def\proof{\noindent{\emph{Proof:} }}

\def
\endproof{\hspace*{\fill}~\qed
\par
\endtrivlist\unskip}
\def\endproof{\hspace*{\fill}~\qed\par\endtrivlist\vskip3pt}

\def\phi{\varphi}

\def\l{\left}
\def\r{\right}
\def\({\left(}
\def\){\right)}

\def\b0{{\mathbf{0}}}

